\documentclass[aps,pra,10pt,superscriptaddress,numerical,floatfix,twocolumn]{revtex4-2}

\usepackage{bm}
\usepackage{braket}
\usepackage{color}
\usepackage{dcolumn}
\usepackage{epsfig}
\usepackage{float}
\usepackage[T1]{fontenc}
\usepackage{graphicx}
\usepackage[colorlinks]{hyperref}
\AtBeginDocument{% optionally set colors to your liking
	\hypersetup{
		urlcolor=black,
		citecolor=green,
		linkcolor=blue,
	}%
}
\usepackage[latin9]{inputenc}
\usepackage{mathtools}
\usepackage{caption}
\usepackage{subcaption}
\usepackage{xcolor}
\usepackage{cleveref}
\usepackage{amssymb}
\usepackage{amsthm}
\usepackage{amsmath}
 
\def\app#1#2{%
	\mathrel{%
		\setbox0=\hbox{$#1\sim$}%
		\setbox2=\hbox{%
			\rlap{\hbox{$#1\propto$}}%
			\lower1.2\ht0\box0%
		}%
		\raise0.25\ht2\box2%
	}%
}

\begin{document}

\title{Unitary Averaging with Fault and Loss Tolerance}

\newcommand{\affone}{Centre for Quantum Computation and Communication Technology, School of Mathematics and Physics, University of Queensland, Brisbane, Queensland 4072, Australia.}

\newcommand{\afftwo}{Dahlem Center for Complex Quantum Systems,	Freie Universit\"at Berlin, 14195 Berlin, Germany}

\author{Ryan J. Marshman}
\affiliation{\affone}

\author{Deepesh Singh}
\affiliation{\affone}

\author{Timothy  C. Ralph}
\affiliation{\affone}

\author{Austin P. Lund}
\affiliation{\afftwo}
\affiliation{\affone}

\date{\today}

\begin{abstract}
	{We consider the impact of the unitary averaging framework on single and two-mode linear optical gates. We demonstrate that this allows a trade-off between the probability of success and gate fidelity, with perfect fidelity gates being achievable for a finite decrease in the probability of success, at least in principle. Furthermore, we show that the encoding and decoding errors in the averaging scheme can also be suppressed up to the first order. We also look at how unitary averaging can work in conjunction with existing error correction schemes. Specifically, we consider how parity encoding might be used to counter the extra loss due to the decreased probability of success, with the aim of achieving fault tolerance. We also consider how unitary averaging might be utilised to expand the parameter space in which fault tolerance may be achievable using standard fault tolerant schemes.
 }

\end{abstract}

%\pacs{}
\maketitle
\section{Introduction}

With noisy, intermediate-scale quantum (NISQ) devices becoming more common the attention is shifting to the creation of full-scale quantum computing devices \cite{nielsen2010quantum}. These future devices will differ to current technology in their scale, which necessitates the development of practical methods of dealing with noise in quantum devices. Specifically, full scale quantum computation will require the used of fault-tolerant schemes. However,  these require an enormous overhead in resources to be implemented. Quantum computing architectures and error-correction schemes that offer a reduction in this overhead will be necessary for at least near future quantum computing devices and is thus the focus of much interest. Indeed, improvements in this space which are easily implementable would be useful both in the NISQ-era and in the longer term.

Linear optics provides a promising platform for achieving the transition from existing technologies like Boson sampling \cite{broome2013photonic} to universal quantum computing realisations, like fusion-based quantum computing \cite{bartolucci2023fusion}. The implementation of deterministic two-qubit entangling gates with non-interacting photons, photon loss, and the build up of errors throughout the quantum circuits represent some of the major challenges for optical quantum computers. The first two issues can be addressed by loss tolerant encoding \cite{ralph2005loss} to enable recovery from failed probabilistic components. As such loss tolerance must be a fundamental component of any optical quantum computer, a fact that our scheme relies on to address the limitation of its probabilistic nature.

We explore the framework of Unitary Averaging (UA) \cite{marshman2018passive, vijayan2020robust, singh2022optical} which allows one to alleviate the effect of imperfections within the applied transformations. UA employs redundancy in the transformation circuit to passively, but probabilistically, reduce the build up of errors. The scheme has a simple implementation within linear optical systems, and so we will consider this as the explicit example system. UA a transformation equivalent to the average of a given set of transformations. In the limit when all the individual transformations are close approximations of some target, their average is a good approximation of the target in total variation distance (TVD). The resulting improvement in the gate fidelity however comes at the cost of success probability, and we analyse it in all generality in the linear optical architecture.  %Deepesh, can you please provide a reference for the TVD part?

As is the case with all error-correction schemes, encoding errors present a challenge to their practical usefulness. An exploration of the same is therefore considered for the UA framework here. It is shown that any errors present in any encoding utilised in the UA scheme, is naturally suppressed to the first order.

 Moreover, the trade-off between gate fidelity and loss may be useful given the necessity of loss protection in any realistic optical quantum computer due to both photon absorption and the probabilitic nature of optical quantum computation \cite{knill2001scheme, rudolph2017optimistic}. We consider the loss-tolerant parity encoding scheme and show its compatibility with UA. Thus, in the limit of large redundancy, UA could be employed to ensure all logical errors are converted into heralded losses which are recoverable by the parity encoding. 
 
 Finally, we consider how UA may be used to expand the parameter space for which fault tolerance can be achieved by analysing a few example error codes. Specifically, we consider some older fault tolerant codes (7-qubit Steane code and the 23-qubit Golay code \cite{dawson2006noise}) for which the benifit of UA is seen to be significant, as well as more modern surface code based implementations \cite{fujii2012error, bartolucci2023fusion} for which effect of UA is more modest.

The structure of this paper is as follows: we begin by summarising the known results on UA in section \ref{sec:UA intro}. Section \ref{sec:UA encoding errors} explores the suppression of encoding errors in single-mode unitary averaged gates while section \ref{sec:UA single qubit gates} details the effect unitary averaging has on an arbitrary single-mode unitary in terms of both the gate fidelity and probability of success. Section \ref{sec:Two qubit gates} discusses how two-mode gates are implemented and protected in a similar fashion. We then introduce the loss tolerant scheme of parity encoding in section \ref{sec:PE intro} before highlighting how it can be used together with UA in section \ref{sec:Combining PE and UA}. Section \ref{sec:FT encodings} then discusses the use of UA with the threshold for fault tolerance before some concluding remarks in section \ref{sec:Conclusion}.

\section{Unitary Averaging \label{sec:UA intro}}
Given access to multiple unitaries, the unitary averaging (UA) framework allows one to apply an average of these unitaries on the intended modes, with the success probability of this transformation depending on the exact value of the unitaries. 

Unitary Averaging (UA) acts to apply an averaged unitary evolution \cite{marshman2018passive, vijayan2020robust} given by
\begin{equation}
	\hat{\mathcal{U}}=\frac{1}{N}\sum_{j=1}^{N}\hat{U}_j
\end{equation}
on a target set of $M$ modes. There is also an accompanying $(N-1)M$ set of error modes which are post-selected to be in a vacuum state. In this way, if each applied unitary $\hat{U}_j$ are approximately implementing a target unitary $\hat{U}_T$ with a certain unbiased noise, then by the central limit theorem, UA will instead apply an averaged unitary $\hat{\mathcal{U}}$ which represents a new, stochastic operator which approximates the target unitary $\hat{U}_{T}$, with variance reduced by a factor of $N$. As demonstrated later, we can further write the individual transformation in the form $\hat{U}_{j}=\hat{U}_{T}+\hat{E}_{j}$ where $\hat{E}_{j}$ is the stochastic operator containing all of the noise terms. Note that the use of $\mathcal{U}$ rather than $U$ is to remind the reader that the resulting transformation is non-unitary, but aims to be as close to unitary as possible. Specifically, the target unitary can be implemented arbitrarily accurately using a sufficiently large $N$. The cost of this reduction in variation of the applied transformation is that it is implemented probabilistically. The probability of success $P_{S}(N)$ depends on both the variance in the individual unitaries, and number of copies used $N$. The decoding produces $N-1$ error heralding modes. Detecting a photon at the error heralding modes applies the transformation
\begin{equation}
	\hat{\mathcal{U}}_{e}=\frac{1}{N}\sum_{j=1}^{N}f_{j}\hat{U}_j
\end{equation}
where the weights $f_{j}$ are real numbers such that $0 \leq f_{j} \leq 1, \forall j$ and $\sum_{j=1}^{N}f_{j} = 1$. The values $f_{i}$ are entirely depending on the encoding and decoding used in the averaging process. When the encoding unitary is $H^{\otimes k}$ for any $k$, $f_j=\pm1$ such that at least one $f_{j}=-1$, as will be used throughout this paper. As such, observing a photon in the heralding modes applies an unintended transformation, furthermore, depending on the nature of the detectors used, it may destroy the state. As such, UA can be viewed as mapping logical errors to heralded loss and phase error imprinted on the remainder of the state. The probability of such a heralded error scales proportionally to the variance of the applied unitaries $\hat{U}_j$ and the amount of averaging, $N$.

Throughout this manuscript we are concerned only with the success modes. As such, we consider the output state to only be the correct modes, dropping the error heralding modes. The output state is thus
\begin{equation}
	\hat{\rho}(N)=\,\hat{\mathcal{U}}(N)\left|\psi\right\rangle\left\langle\psi\right|\hat{\mathcal{U}}^{\dagger}(N)
\end{equation}
which is an unnormalised state, and where $\left|\psi\right\rangle$ is the initial state. The normalisation is returned though the post selection process, which is then the source of the probabilistic nature of the correction being applied. This is characterised by the probability of success $P_{s}(N)$. In fact, we can define this the probability of success by the amplitude of the un-normalised state which in this case can be calculated for a general input state using
\begin{align}
	P_{s}(N)=&\left|\hat{\mathcal{U}}(N)\left|\psi\right\rangle\right|^2  \nonumber\\
	%P_{s}(N)=&\left\langle \psi\right| \hat{U}^{\dagger}(N)\hat{U}(N)\left|\psi\right\rangle\nonumber\\
	=&\frac{1}{N^2}\sum_{j=1}^{N}\sum_{k=1}^{N}\left\langle \psi\right| \hat{U}_{j}^{\dagger}\hat{U}_{k}\left|\psi\right\rangle \label{eq:Prob success definition} 
\end{align}

The density operator after the post selection step is then given by
\begin{equation}
	\hat{\rho}_{ps}=\left(P_s(N)\right)^{-1}\hat{\rho}(N). \label{eq:post selected density operator}
\end{equation}

The other figure of merit to characterise the effect UA has on a gate is the state fidelity. The fidelity encodes how likely it is that, upon measurement, the transformation returns the target state. The target state in this case is $\left|\Psi\right\rangle=\hat{U}_{T}\left|\psi\right\rangle$ where $\left|\psi\right\rangle$ is the initial state and $\hat{U}_{T}$ is the target unitary, which corresponds to $\hat{U}_{j}$ in the instance of no noise. The gate fidelity is then defined as
\begin{align}
	\mathcal{F}(N)=&\left\langle \Psi\right| \rho_{ps}(N)\left|\Psi\right\rangle\label{eq:fidelity definition}
\end{align}

Before we can start characterising the effect UA has on single and two qubit gates, we must also consider the encoding and decoding steps and to what extent they impact on the output.

\section{unitary averaging Encoding Errors \label{sec:UA encoding errors}}

\begin{figure}[h]
	\centering
	\begin{subfigure}[l]{0.9\columnwidth}
		\includegraphics[width=\columnwidth]{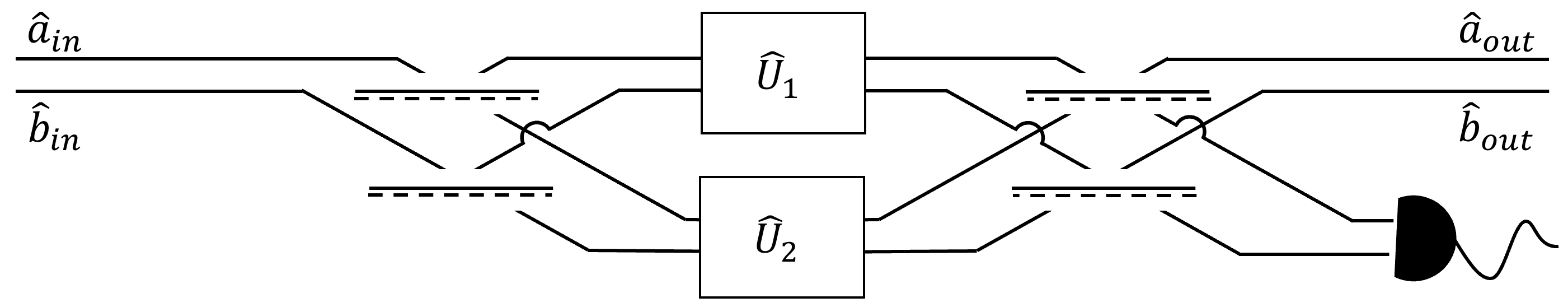}
		\caption{$N=2$ \label{fig:n=2 UA}}
	\end{subfigure}
	\begin{subfigure}[l]{0.9\columnwidth}
		\includegraphics[width=\columnwidth]{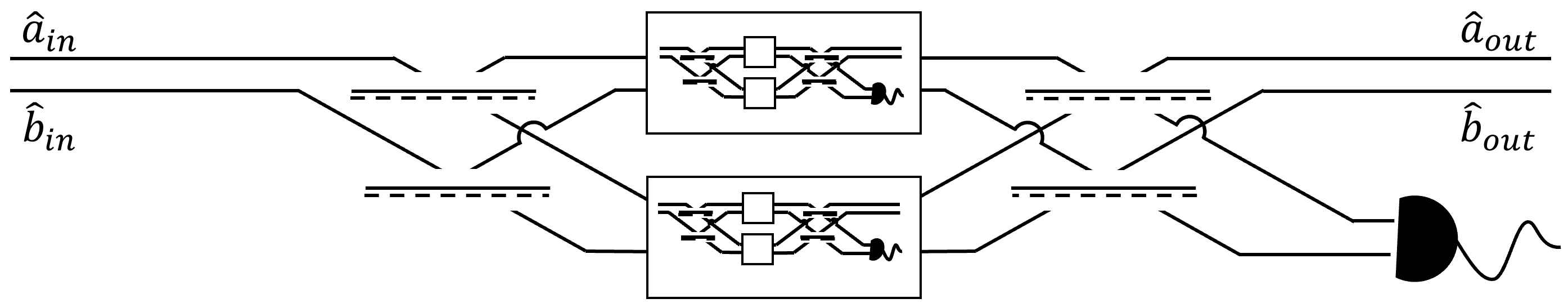}
		\caption{$N=4$ \label{fig:n=4 UA}}
	\end{subfigure}
	\caption{Arbitrary dual-rail, single qubit unitary with unitary averaging. Figure \ref{fig:n=2 UA} shows $N=2$ levels of redundancy while Figure \ref{fig:n=4 UA} shows how $N=4$ can be viewed similarly, but where each physical unitary $\hat{U}_{1/2}$ as itself an $N=2$ encoded unitary. \label{fig:N=2 vs N=4}}
\end{figure}

The encoding and decoding can be achieved in multiple different manners, including using Hadamard encoding~\cite{marshman2018passive}, or the more general W-state/ quantum fourier transform encodings~\cite{vijayan2020robust, singh2022optical}. In this instance we will consider the Hadamard encoding which is simple in its construction and in particular its scaling to higher levels of encoding ($N$). Specifically, we will take the single qubit gates to be encoded using only beam splitters, splitting the input evenly between each redundant physical gate. This can then be performed iteratively, giving $N=2^n$ for $n\in\mathbb{N}$. This process is shown in Figure \ref{fig:N=2 vs N=4}. This choice of encoding is also optimal in that it maintains constant optical depth of each interferometric path while minimising the optical depth. Given the even splitting, the optical depth for each path increases logarithmically as $2\log_{2}\left(N\right)$ for $N$ copies of the unitary.

Starting with $N=2$, the output of a single qubit averaged gate after post selection is determined by 
\begin{equation}
	\left|\psi\right\rangle_{out}=\left\langle\bf{0}\right|_{\epsilon}\hat{B}_{1\xleftrightarrow[]{}2}^{(2)}\left(\hat{U}_2\otimes\hat{U}_1\right)\hat{B}_{1\xleftrightarrow[]{}2}^{(1)}\left|\psi\right\rangle_{in}. \label{eq:N=2 general single qubit transformation}
\end{equation}
where $\hat{B}_{a\xleftrightarrow[]{}b}^{(j)}$ acts to evenly mix the pair of modes $a$ and $b$, each of which are then separately acted on by independent unitary $\hat{U}_{a/b}$ and $\left\langle\bf{0}\right|_{\epsilon}$ represents the projection onto the vacuum for the error modes. The superscript $j$ serves to remind that each beam splitter is independent and thus has its own unique noise associated with it. In dual rail, this can be written as
\begin{equation}
	\hat{B}_{1\xleftrightarrow[]{}2}^{(j)}=\begin{bmatrix}
		 \sin\left(\theta_j\right) & 0  & \cos\left(\theta_j\right) & 0 \\
		 0 & \sin\left(\theta_j'\right)  &0 & \cos\left(\theta_j'\right)  \\
		\cos\left(\theta_j\right) & 0 & -\sin\left(\theta_j\right) & 0  \\
		 0 & \cos\left(\theta_j'\right)  & 0 & -\sin\left(\theta_j'\right)
	\end{bmatrix}
\end{equation}
where  $\theta_{j}^{(')}=\frac{\pi}{2}+\delta\theta_j^{(')}$ for $\delta\theta_j^{(')}\ll\frac{\pi}{2}$, allowing us to write $\sin\left(\theta_j\right)\approx\frac{1}{\sqrt{2}} + \delta\theta_j^{(')}$ and $\cos\left(\theta_j\right)\approx\frac{1}{\sqrt{2}} - \delta\theta_j^{(')}$.  For $N=4$, given the concatinated nature of the encoding choice, single qubit gate $\hat{U}_{1/2}$ can be be  replaced with a $N=2$ circuit ($\hat{B}_{1\xleftrightarrow[]{}2}^{(2)}\left(\hat{U}_2\otimes\hat{U}_1\right)\hat{B}_{1\xleftrightarrow[]{}2}^{(1)}$) and relabelling the elements. This process can then be further repeated for higher $N=2^{j}$ where $j\in\mathbb{N}$. However, care is needed when relabelling each single qubit unitary and beam splitter to ensure each act on the appropriate modes.

To calculate the transformation implemented in Equation \ref{eq:N=2 general single qubit transformation} after heralding, but before renormalisation on no error detection results in
\begin{widetext}
\begin{align}
	\hat{a}_{out}\left(N=2\right)=&\left[\sin\left(\theta_{1}\right)\sin\left(\theta_{2}\right)\hat{U}_1+\cos\left(\theta_{1}\right)\cos\left(\theta_{2}\right)\hat{U}_2\right]\hat{a}_{in}+\left[\sin\left(\theta_{1}'\right)\sin\left(\theta_{2}'\right)\hat{U}_1+\cos\left(\theta_{1}'\right)\cos\left(\theta_{2}'\right)\hat{U}_2\right]\hat{b}_{in} \nonumber\\
	\approx&\frac{1}{2}\left(\hat{U}_1 + \hat{U}_2\right)\left(\hat{a}_{in}+\hat{b}_{in}\right)+\frac{1}{\sqrt{2}}\left(\hat{U}_1 -\hat{U}_2 \right)\left[\left(\delta\theta_1+\delta\theta_2\right)\hat{a}_{in}+\left(\delta\theta_1'+\delta\theta_2'\right)\hat{b}_{in}\right]
\label{eq:n=2 averaging encoding errors 1}
\end{align}
%\begin{align}
%	\hat{U}_{UA}\left(N=2\right)=&\sin\left(\theta_{1}\right)\sin\left(\theta_{2}\right)\hat{U}_1+\cos\left(\theta_{1}\right)\cos\left(\theta_{2}\right)\hat{U}_2 \nonumber\\
%	\approx&\frac{1}{2}\left(\hat{U}_1 + \hat{U}_2\right)+\frac{1}{\sqrt{2}}\left(\delta\theta_1+\delta\theta_2\right)\hat{U}_1 \nonumber\\
  %  & - \frac{1}{\sqrt{2}}\left(\delta\theta_1+\delta\theta_2\right)\hat{U}_2 \label{eq:n=2 averaging encoding errors 1}
%\end{align}
which can be simplified by taylor expanding the applied single qubit gates, setting $\hat{U}_{j}=\hat{U}_{T}+\hat{E}_{j}$ (as discussed in more detail later), and keeping only linear terms simplifies to
\begin{align}
	\hat{a}_{out}\left(N=2\right)\approx&\left(\hat{U}_T+\frac{1}{2}\left(\hat{E}_1 + \hat{E}_2\right)\right)\left(\hat{a}_{in}+\hat{b}_{in}\right)+\mathcal{O}\left(\hat{E}^2,~\delta\theta\hat{E},~\delta\theta^2\right). \label{eq:n=2 averaging encoding errors 2}
\end{align}
\end{widetext}
with all terms linear in $\delta\theta_{j}$ cancelling. An equivalent expression can be written for $\hat{b}_{out}$ .
Similarly, the $N=4$ result can be calculated by substituting each single qubit unitary $\hat{U}_i$ with the entire $N=2$ result, with the appropriate re-labelling of the parameters giving
%\begin{widetext}
%\begin{align}
%	\hat{U}_{UA}\left(N=4\right)\approx&\frac{1}{4}\left(\hat{U}_1 + \hat{U}_2 + \hat{U}_3 + \hat{U}_4\right) + \frac{1}{2\sqrt{2}}\left(\delta\theta_1+\delta\theta_2+\delta\theta_3+\delta\theta_4\right)\hat{U}_1 + \frac{1}{2\sqrt{2}}\left(\delta\theta_1+\delta\theta_2-\delta\theta_3-\delta\theta_4\right)\hat{U}_2\nonumber\\ &-\frac{1}{2\sqrt{2}}\left(\delta\theta_1+\delta\theta_2-\delta\theta_5-\delta\theta_6\right)\hat{U}_3 - \frac{1}{2\sqrt{2}}\left(\delta\theta_1+\delta\theta_2+\delta\theta_5+\delta\theta_6\right)\hat{U}_4 \nonumber\\
%	\approx&\hat{U}+\frac{1}{4}\left(\hat{E}_1 + \hat{E}_2 + \hat{E}_3 + \hat{E}_4\right) +\mathcal{O}\left(\hat{E}^2,~\delta\theta\hat{E},~\delta\theta^2\right)
%\end{align}
%\end{widetext}
\begin{align}
	\hat{a}_{out}\left(N=4\right)\approx&\left(\hat{U}_T+\frac{1}{4}\Sigma_{j=1}^{4}\hat{E}_{j}\right)\left(\hat{a}_{in}+\hat{b}_{in}\right) \nonumber\\
 &+\mathcal{O}\left(\hat{E}^2,~\delta\theta\hat{E},~\delta\theta^2\right)
\end{align}
where again the linear encoding error terms cancel one another. One can see by induction that this pattern continues for all $N=2^{i}$ where $i\in\mathbb{Z}$.  Thus encoding errors are naturally suppressed to the first order.  As such, it is sufficient to account for only errors within the averaged unitaries themselves so encoding errors will not be considered for the remainder of this manuscript.

One might be tempted to view this suppression of the encoding errors as a result of the concatenation, however, given the linear encoding errors for the $N=2$ case (as shown in Eq. \ref{eq:n=2 averaging encoding errors 2}) this cannot be the full story. While the concatenation will help further suppress noise, it is the post selection which removes the last two terms from Eq. \ref{eq:n=2 averaging encoding errors 2} combined with the correlated noise in the encoding resulting here from the use of a single beam splitter for both polarisation modes. Maintaining this benefit with multiple qubits will require the noise to again be correlated. This could only be achieved by considering the implementation specificially. For example, using a single thermal coupling controller to manipulate both beam-paths if thermal control is used to modify the distance and hence evanescent coupling strength between two modes may be sufficient. 

\section{unitary averaged single qubit gates \label{sec:UA single qubit gates}}
Here we consider how an arbitrary single physical qubit gate might be protected using UA. Specifically we detail how phase and bit-flip errors are converted into heralded loss allowing for a trade-off between fidelity and known loss. Given the results of the previous section, it is not necessary to include any encoding errors and so we will consider noise only in the linear optical unitary.

\begin{figure}[h]
	\centering
	\includegraphics[width=0.8\columnwidth]{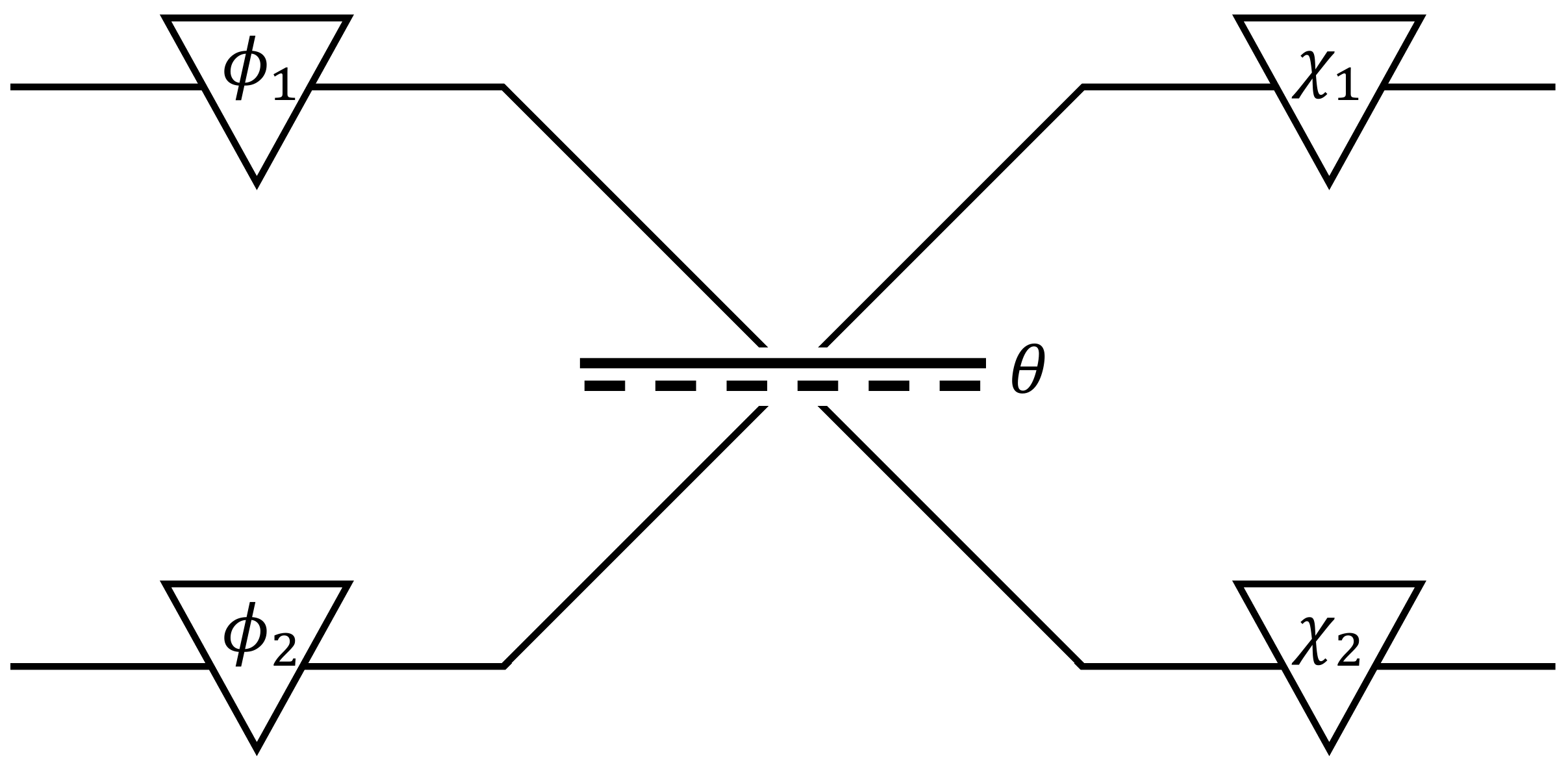}
	\caption{Arbitrary dual-rail, single qubit unitary. \label{fig:arbitrary single qubit unitary}}
\end{figure}

The first thing to do is build an error model for an unencoded gate and then considering the impact of UA on these errors. To do this we will use an over complete gate description so as to allow errors to arise anywhere within the unitary. The circuit depth is also then the same regardless of path taken. The chosen system allows an arbitrary single qubit gate to be implemented, with specific parameters tuned such that it implements the intended transformation. The system is shown in Figure \ref{fig:arbitrary single qubit unitary}.
This gate will implement the transformation
\begin{equation}
	\hat{U}_{j}=\begin{bmatrix}
		e^{i\phi_{1,j}}e^{i\chi_{1,j}}\sin\left(\theta_{j}\right) & e^{i\phi_{2,j}}e^{i\chi_{1,j}}\cos\left(\theta_{j}\right) \\
		e^{i\phi_{1,j}}e^{i\chi_{2,j}}\cos\left(\theta_{j}\right) & -e^{i\phi_{2,j}}e^{i\chi_{2,j}}\sin\left(\theta_{j}\right)
	\end{bmatrix}
\end{equation}
We can then take each parameter $O_{i,j}$ to be given by the intended target value $O_{i}$ and an additional noise term $\delta O_{i,j}$ where $O_{i,j}=O_{i}+\delta O_{i,j}$. The appropriate gate parameters for a number of unitaries of interest are shown in Table \ref{tab:signle qubit gate parameters}.
\begin{table}
\begin{center}
\begin{tabular}{|c||c|c|c|c|c|}
	\hline
	Gate& $\theta$ & $\phi_{1}$ & $\phi_{2}$ & $\chi_{1}$ & $\chi_{2}$ \\ [0.5ex]
	\hline
	\hline
	$\hat{I}$ & $\frac{\pi}{2}$ & $0$ & $0$ & $0$ & $\pi$ \\
	$\hat{X}$ & $0$ & $0$ & $0$ & $0$ & $0$ \\
	$\hat{Y}$ & $0$ & $\frac{\pi}{2}$ & $0$ & $-\frac{\pi}{2}$ & $0$ \\
	$\hat{Z}_{\alpha}$ & $\frac{\pi}{2}$ & $0$ & $0$ & $0$ & $\alpha$ \\
	$\hat{H}$ & $\frac{\pi}{4}$ & $0$ & $0$ & $0$ & $0$ \\
	\hline
\end{tabular}
\end{center}
\caption{Arbitrary single qubit gate parameters \label{tab:signle qubit gate parameters}.}
\end{table}
We then Taylor expand each parameter around their target value as shown in the supplemental information for this paper. The applied unitary can in general be written as $\hat{U}_{j}=\hat{U}_{T}+\hat{E}_{j}$ where $\hat{U}_{T}$ is the target unitary. Thus, after employing UA, we will have implemented the transformation on the output mode 
\begin{equation}
	\hat{\mathcal{U}}(N)=\frac{1}{N}\sum_{j=1}^{N}\left(\hat{U}_{T}+\hat{E}_{j}\right)\equiv\hat{U}_{T}+\frac{\hat{\epsilon}(N)}{N}
\end{equation}
where $\hat{\epsilon}(N)$ is some stochastically varying transformation and we are not tracking what happens to the error modes as we will post-select these to be in the vacuum. Thus, so long as each error is unbiased and independent, we might expect any desired gate fidelity to be achievable with sufficiently large $N$. Specifically with $\lim_{N\rightarrow\infty}\hat{\mathcal{U}}(N)=\hat{U}_{T}$, i.e. arbitrarily high gate fidelity may be possible with sufficient redundancy. However, this will be at the cost of a decreasing probability of success, $P_{s}(N)$. 

As such, we can write the general, unnormalised state at the output modes (i.e. after post selection but without renormalising the state) as
\begin{align}
	\hat{\rho}(N)=&\,\hat{\mathcal{U}}(N)\left|\psi\right\rangle\left\langle\psi\right\rangle\hat{\mathcal{U}}^{\dagger}(N)\nonumber\\
	=&\left(\hat{U}_{T}+\frac{\hat{\epsilon}(N)}{N}\right)\left|\psi\right\rangle\left\langle\psi\right|\left(\hat{U}_{T}+\frac{\hat{\epsilon}^{\dagger}(N)}{N}\right) \nonumber\\
	=&\left|\Psi\right\rangle\left\langle\Psi\right| + \frac{1}{N}\left(\hat{\epsilon}(N)\left|\psi\right\rangle\left\langle\psi\right|\hat{U}_{T} + \hat{U}_{T}\left|\psi\right\rangle\left\langle\psi\right|\hat{\epsilon}^{\dagger}(N)\right)\nonumber\\
	& + \frac{1}{N^2}\hat{\epsilon}(N)\left|\psi\right\rangle\left\langle\psi\right|\hat{\epsilon}^{\dagger}(N)
\end{align}
and the density operator after post selection is again given by
\begin{equation}
	\hat{\rho}_{ps}=\left(P_s(N)\right)^{-1}\hat{\rho}(N). \label{eq:post selected density operator}
\end{equation}
with the associated probability of success as define in Equation \ref{eq:Prob success definition}.

The probability of success can here be explicitly solved as determined by the moments of the noise distributions under the assumptions that each noise term is independent. For simplicity we take $\left\langle\delta O\right\rangle=0$, $\left\langle\delta O^2\right\rangle=\nu$, $\left\langle\delta O^3\right\rangle=0$ and $\left\langle\delta O^4\right\rangle=3\nu^2$,  for all noise terms $\delta O$ to give
\begin{equation}
	P_{s}\left(N\right)\approx1-3\nu+\frac{3\nu}{N}+\frac{9\nu^2}{2}-\frac{9\nu^2}{2N}  \label{eq:Psuccess 1}
\end{equation}
which is shown in Figure \ref{fig:PsuccessScaling1}. For the details of this calculation up to first order in $\nu$, see the supplemental information for this paper, the higher order calculation was performed with the aid of \emph{Mathematica}. We have taken the noise to be Gaussian to simplify the result, however this could be replaced by other unbiased probability distributions without significantly impacting the results.
\begin{figure}[h]
	\centering
	\includegraphics[width=0.99\columnwidth]{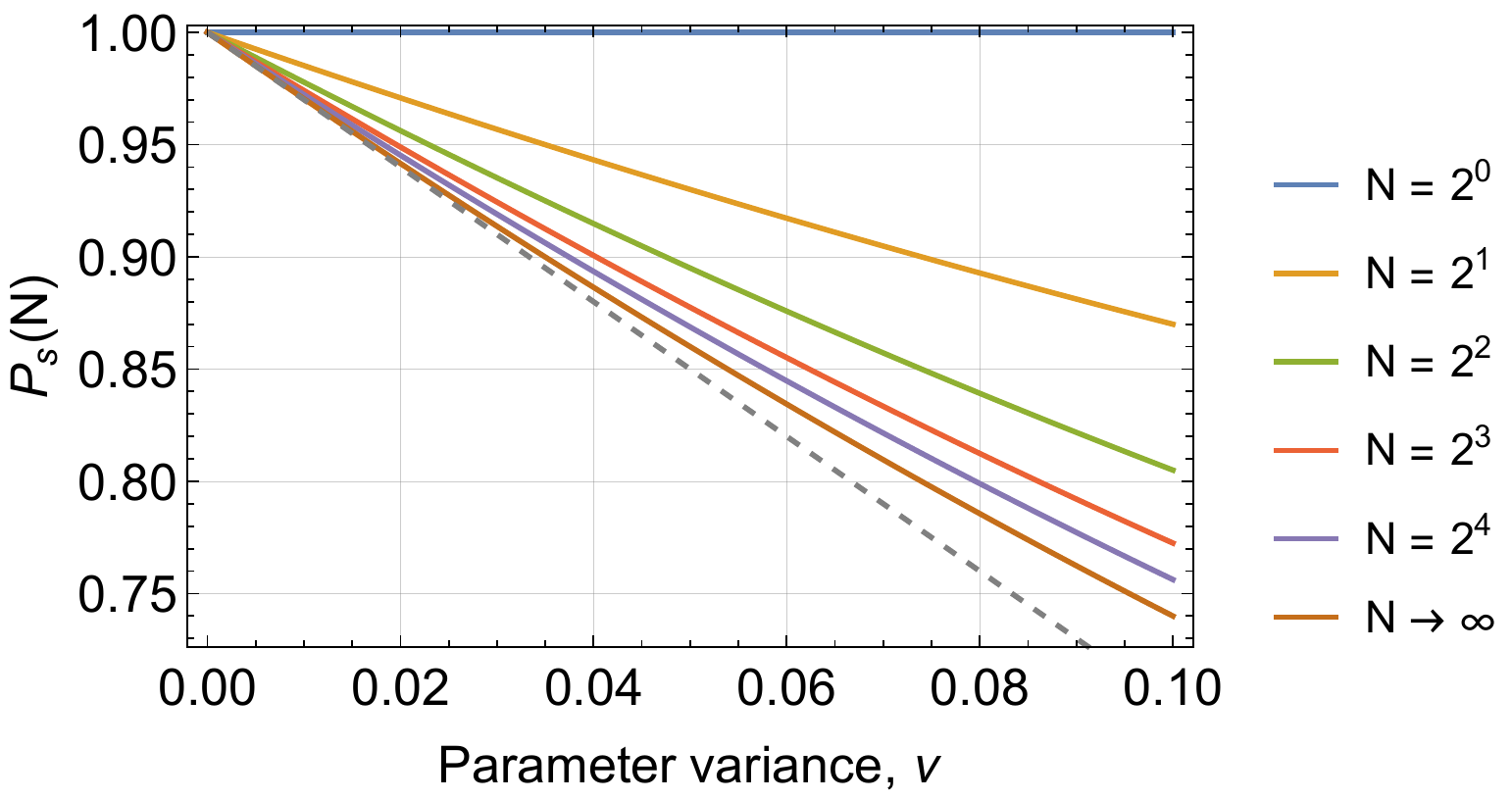}
	\caption{Probability of success scaling for an unitary averaged single qubit gate as given by  Equation \ref{eq:Psuccess 1}. The dashed grey line shows the first order scaling of $\lim_{N\rightarrow\infty}P_{s}\left(N\right)=1-3\nu$. \label{fig:PsuccessScaling1}}
\end{figure}

The post selected gate fidelity can be calculated similarly with
\begin{align}
	\mathcal{F}(N)=&\left\langle \Psi\right| \rho_{ps}(N)\left|\Psi\right\rangle\nonumber\\
	=&\left(P_s(N)\right)^{-1}\left\langle \psi\right| \hat{U}_{T}^{\dagger}\,\hat{\mathcal{U}}(N)\left|\psi\right\rangle\left\langle \psi\right| \hat{\mathcal{U}}^{\dagger}(N)\hat{U}_{T}\left|\psi\right\rangle\nonumber\\
	=&\left(P_s(N)\right)^{-1}\left(1+\frac{1}{N}\sum_{j=1}^{N}\left\langle \psi\right| \hat{U}_{T}^{\dagger}\hat{E}_{j}\left|\psi\right\rangle\right)^2
\end{align}

\begin{figure}[h]
	\centering
	\includegraphics[width=0.99\columnwidth]{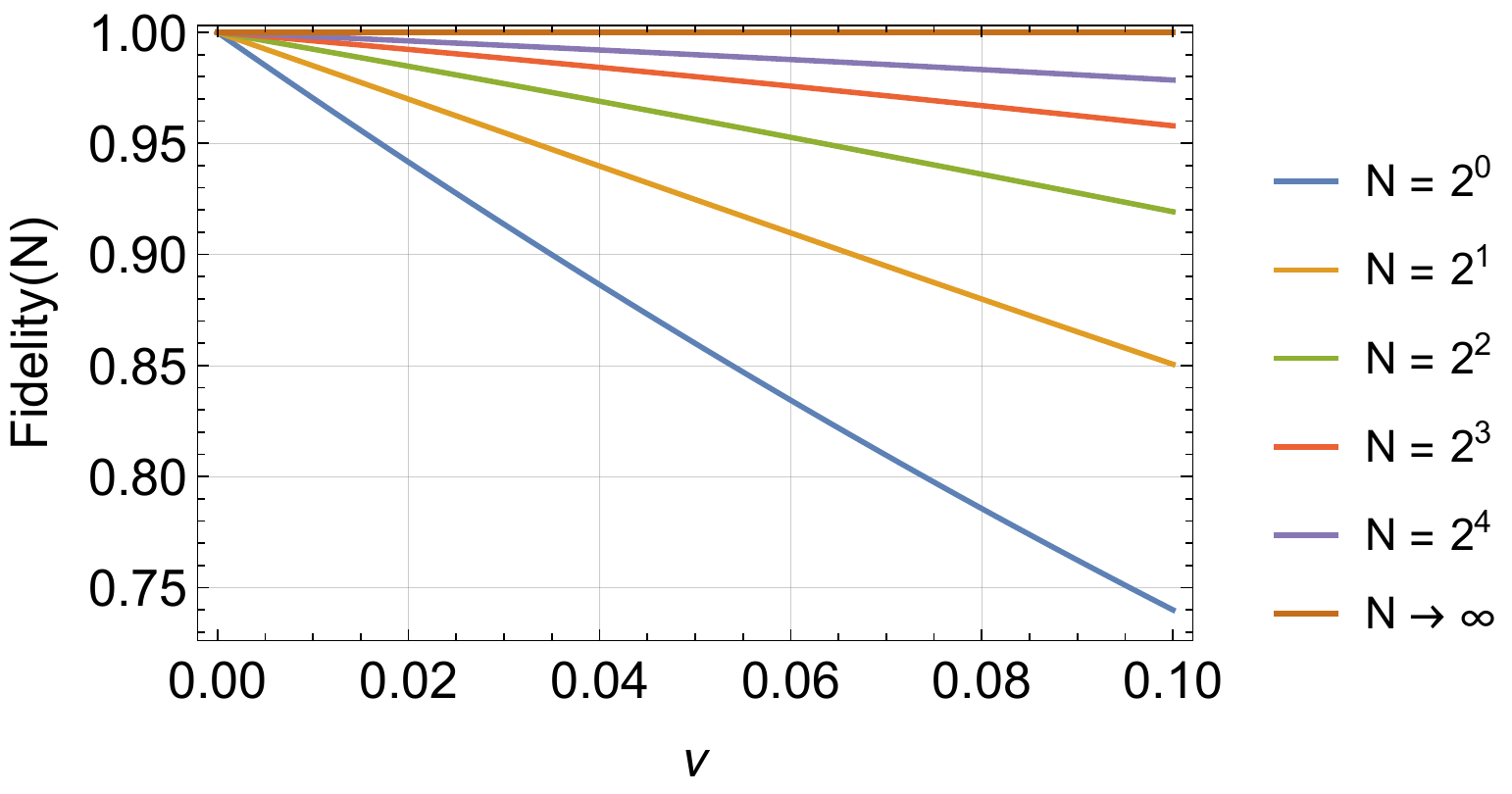}
	\caption{Post-selected Fidelity for an unitary averaged single qubit gate as given by Equation \ref{eq:Fidelity}. \label{fig:Fidelity}}
\end{figure}

Again, the fidelity is characterised by the moments of the noise probability distribution with 
\begin{equation}
	\mathcal{F}\left(N\right)\approx\frac{1-3\nu+\frac{9}{4\nu^2}}{1-3\nu+\frac{\nu}{N}+\frac{9\nu^2}{2}-\frac{9\nu^2}{2N}} \label{eq:Fidelity}
\end{equation}
to the second order in the parameter variance $\nu$. The details of the calculation are provided in the supplemental information. The fidelity scaling is shown in Figure \ref{fig:Fidelity}.

\section{Two Qubit Gates \label{sec:Two qubit gates}}

\begin{figure}
	\centering
     \begin{subfigure}{0.9\columnwidth}
		\includegraphics[width=\columnwidth]{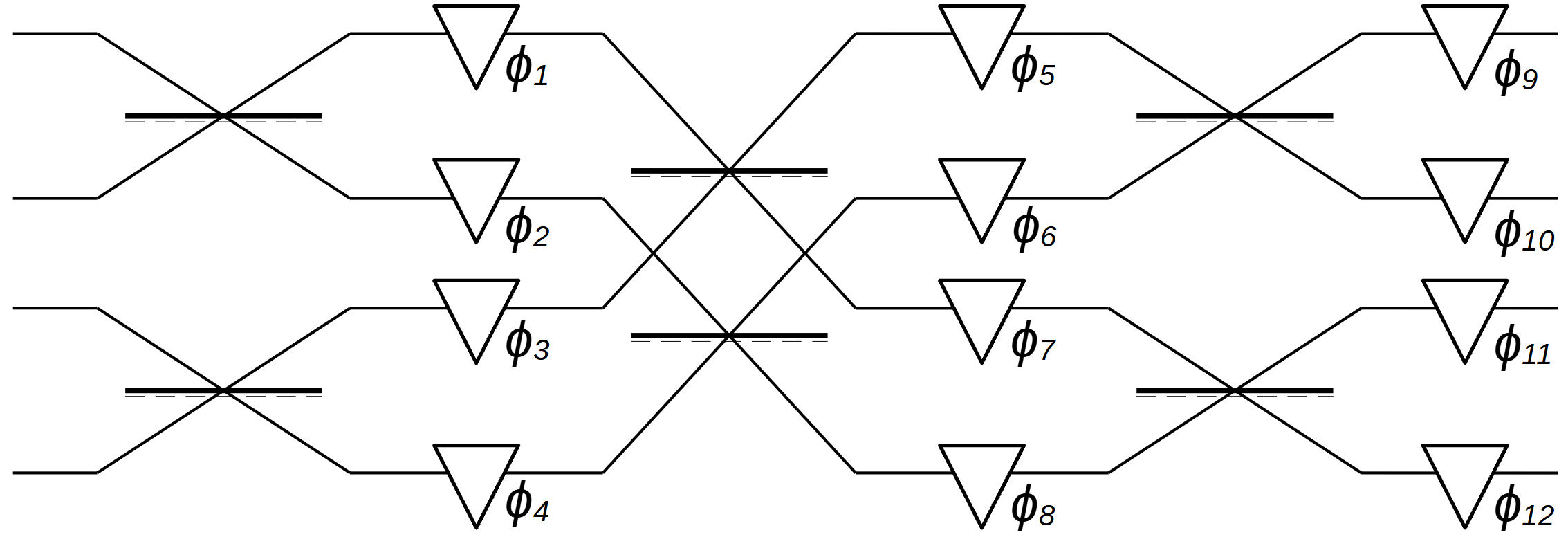}
		\caption{General four mode interferometer.\label{fig:general 4 mode interferometer}}
	\end{subfigure}
     \begin{subfigure}{0.5\columnwidth}
		\includegraphics[width=\columnwidth]{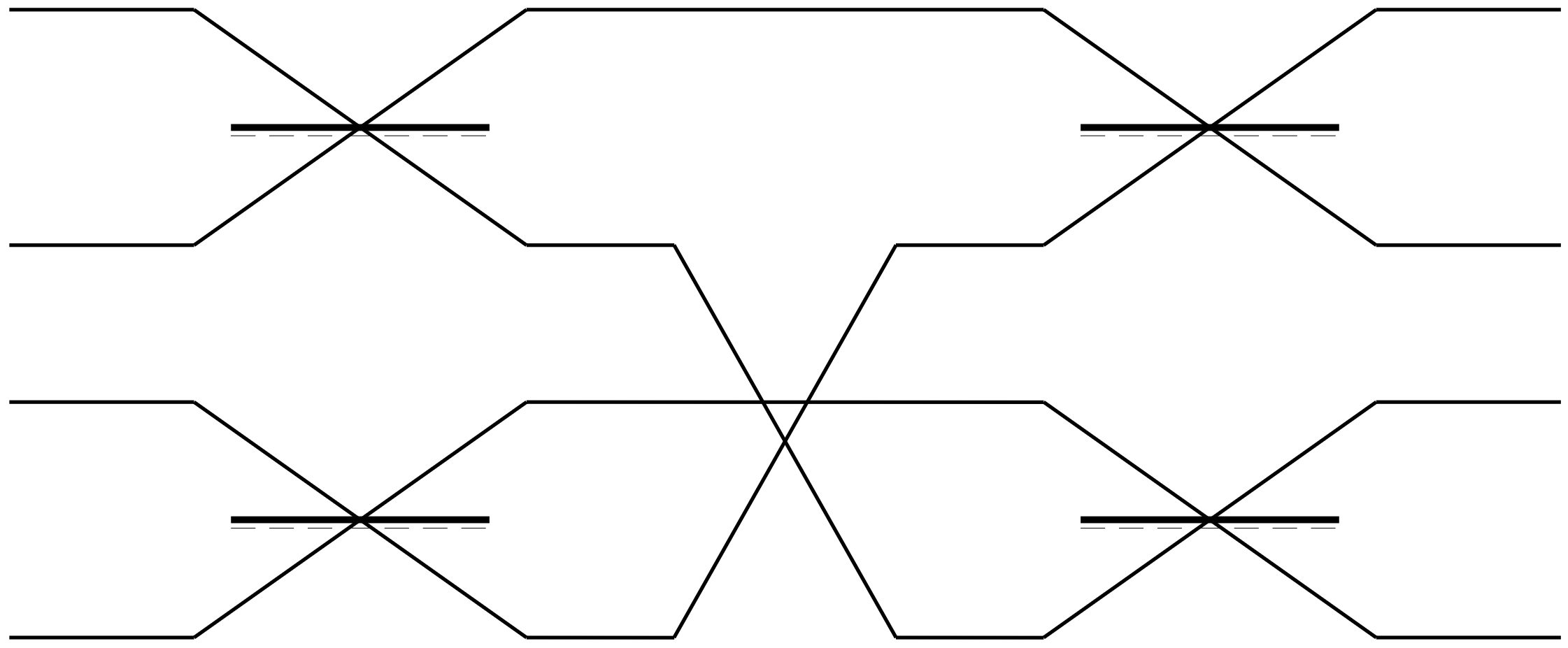}
		\caption{Reduced four mode interferometer capable of acting as a Type-II fusion gate.\label{fig:fusion gate}}
	\end{subfigure}
	\caption{\ref{fig:general 4 mode interferometer} General four mode linear optical gate for two qubits consisting of beam splitters between pairs of modes and phase shifters marked by their phase $\phi_i$. The Type-II fusion can be implemented with a reduced version of this as shown in Figure \ref{fig:fusion gate} with all present beam splitters set to $50:50$. \label{fig:4-mode interferometer}}
\end{figure}

To be sufficient for arbitrary quantum computation it is necessary to have a corrected universal gate set. While we have just shown it is possible to protect any single qubit gate it is still necessary to provide for two qubit gates such as the CNOT gate. Two qubit gates under parity encoding (CF Section \ref{sec:Combining PE and UA}) can be enacted via single qubit gates and re-encoding using type-I and type-II fusion gates \cite{hayes2010fault}. As such, to protect against errors in two qubit gates using unitary averaging, we need to consider how to protect the fusion gates. Recent work has shown that UA can be employed to correct fusion gates \cite{singh2022optical}, however, this study was considering the low $N$ limit. A general two, dual-rail qubit linear optical gate can be implemented by the interferometer diagram shown in Figure \ref{fig:4-mode interferometer}. The fusion gates can be implemented by a simplified version of this system, with additional measurements on the outputs which are taken to occur separately to the UA protocol. 

The same procedure for determining the probability of success for the one photon gate can be employed, only due to the increased problem size \emph{Mathematica} was used to simplify the expressions. It was found that the four mode interferometer shown in Figure \ref{fig:4-mode interferometer}, when employing UA, will implement the target gate with a post selected fidelity of
\begin{equation}
	\mathcal{F}_{\textrm{4-mode}}(N)=P_{\textrm{s, 4-mode}}(N)^{-1}\left(1-6\nu\right) \label{eq:ps fidelity 4 mode}
\end{equation}
and a success probability
\begin{equation}
	P_{\textrm{s, 4-mode}}(N)=1-6\nu+\frac{6\nu}{N}+18\nu^2-\frac{18\nu^2}{N^2}. \label{eq:ps psuccess 4 mode}
\end{equation}
These can be further improved in systems which do not require the full generality enabled by such an interferometer. Specifically, when implementing Type-II fusion gate, no phase manipulation is required, and the central column of beam splitters can be replaced with a single, physical swapping of the second and fourth modes. With this reduced system, the post selected fidelity becomes
\begin{equation}
	\mathcal{F}_{\textrm{II}}(N)=P_{\textrm{s, II}}(N)^{-1}\left(1-2\nu+2\nu^2\right)\label{eq:ps fidelity type2 fusion}
\end{equation}
with a success probability of 
\begin{equation}
	P_{\textrm{s, 4-mode}}(N)=1-2\nu+\frac{2\nu}{N}+2\nu^2-\frac{2\nu^2}{N}.\label{eq:ps psuccess type2 fusion}
\end{equation} 

It appears that the gate fidelity and probability of success scale linearly with the optical depth of the circuit, as seen by comparing Equations \ref{eq:Fidelity}, \ref{eq:ps fidelity 4 mode}, and \ref{eq:ps fidelity type2 fusion} and Equations \ref{eq:Psuccess 1}, \ref{eq:ps psuccess 4 mode}, and \ref{eq:ps psuccess type2 fusion}. We can in general write the probability of success and fidelity to the first order in the variance as
\begin{align}
	P_{\textrm{s, g}}(N)=&1-\mathcal{V}+\frac{\mathcal{V}}{N} \label{eq:first order Psuccess} \\
	\mathcal{F}_{g}(N)=&P_{\textrm{s, g}}(N)^{-1}\left(1-\mathcal{V}\right) \nonumber\\
	=& 1-\frac{\mathcal{V}}{N+\mathcal{V}-N\mathcal{V}}\label{eq:first order fidelity}
\end{align}
where $\mathcal{V}=d\times\nu$ is the characteristic noise parameter for a circuit with an optical depth $d$. 

This all suggests that, at the cost of heralded loss, we could achieve arbitrarily high gate fidelities in linear optics with sufficiently large $N$. An obvious hypothetical use case is then employing effectively infinite averaging ($N\rightarrow\infty$) along with loss tolerance to achieve a regime of \emph{effective} fault tolerance. By which we mean that, within the bounds of the error model considered here, each gate implements a logical transformation with perfect fidelity and so errors do not accumulate. However, perfect fidelity only truly occurs for infinite averaging ($N=\infty$) and so may not reasonably constitute true fault tolerance. None the less, the next section presents how this process with the behaviour of UA within standard fault tolerant protocols presenting in Section \ref{sec:FT encodings}.

\section{Parity Encoding \label{sec:PE intro}}

Parity encoding (PE) provides a loss tollerant encoding in which extra physical qubits can be added throughout the computation in responce to the occurance of loss \cite{ralph2005loss, hayes2008loss}. Given that UA turns logical errors into heralded loss, and PE corrects for photon loss, we will here seek to combine the two to produce a general error correction scheme.

Full parity encoding employs two separate encoding steps, parity (P) type and redundancy (R) type. P type encodes logical states as
\begin{align}
	\left|0\right\rangle_{L}=&\left|0\right\rangle^{\left(n\right)}=\left(\left|+\right\rangle^{\otimes n}+\left|-\right\rangle^{\otimes n}\right)/\sqrt{2} \nonumber\\
	\left|1\right\rangle_{L}=&\left|1\right\rangle^{\left(n\right)}=\left(\left|+\right\rangle^{\otimes n}-\left|-\right\rangle^{\otimes n}\right)/\sqrt{2}
\end{align}
where $\left|\pm\right\rangle=\frac{1}{\sqrt{2}}\left(\left|H\right\rangle\pm\left|V\right\rangle\right)$ as we are considering here polarisation encoded qubits. The R type encoding uses $q$ copies of each of these P encoded states. With this, an arbitrary single qubit state is encoded as
\begin{equation}
	\left|\Psi\right\rangle_{L}=\alpha\bigotimes^{q}\left|0\right\rangle^{\left(n\right)} +\beta\bigotimes^{q}\left|1\right\rangle^{\left(n\right)}
\end{equation}
Parity encoding in optical quantum computing has been shown to contain a universal gate set~\cite{hayes2008loss} with $\hat{X}$ and $\hat{Z}$ rotations implemented simply by applying the gate to a single or all physical qubits respectively while other gates require reencoding.

\section{Parity Encoded and unitary averaged Single Qubit Gates \label{sec:Combining PE and UA}}
In this section we explore how parity encoding can be utilised in conjuction with unitary averaging to implement high fidelity logical operations with a high probability of success using only noisy unitary operations. To begin, we present the case in which $J$ errors occur, but only within a single redundant encoded state.
Consider the initial state
\begin{equation}
	\left|\psi\right\rangle_L=\alpha\left|0\right\rangle^{(n)}\bigotimes^{q-1}\left|0\right\rangle^{(n)}+\beta\left|1\right\rangle^{(n)}\bigotimes^{q-1}\left|1\right\rangle^{(n)}
\end{equation}
on which we act a single logical qubit unitary $\hat{U}=\bigotimes_{j=1}^{2n}\hat{u}_{j}$ which has some effect on each individual, physical qubit. Note that here we have singled out a single redundant copy of the parity encoding which we will consider to be the location of the $J<n$ errors. We will take the individual unitaries to be \emph{unitary averaged} with $N\rightarrow\infty$ and the error ports monitored such that the resulting transformation is projected onto either the \emph{correct} target result, or a \emph{heralded loss}. When a heralded loss occurs a unique stochastic phase factor will then be written on to the physical qubit. Thus, unitary averaging acts to transform each physical qubit according to
\begin{align}
	\left|H\right\rangle_{i}\rightarrow& \hat{u}_{T}\left|H\right\rangle_{i}+\sum_{k=2}^{N}\delta H_{i,k}\left|\epsilon_{k}\right\rangle_{i} \\
	\left|V\right\rangle_{i}\rightarrow& \hat{u}_{T}\left|V\right\rangle_{i}-\sum_{k=2}^{N}\delta V_{i,k}\left|\epsilon_{k}\right\rangle_{i}
\end{align}
where we have neglected normalisation as after projecting on either the error or no error result, it will not be important. Here each $\delta H_{i,k}$ and $\delta V_{i,k}$ is the unique, stochastic phase factor and $\left|\epsilon_{k}\right\rangle_{i}$ represents the photon in the $k$th error mode which are to be measured, heralding a photon loss. 

Any parity encoded state can be expanded, using the identity
\begin{align}
	\left|0\right\rangle^{(j+k)}=&\frac{1}{\sqrt{2}}\left(\left|0\right\rangle^{(j)}\left|0\right\rangle^{(k)}+\left|1\right\rangle^{(j)}\left|1\right\rangle^{(k)}\right)\\
	\left|1\right\rangle^{(j+k)}=&\frac{1}{\sqrt{2}}\left(\left|0\right\rangle^{(j)}\left|1\right\rangle^{(k)}+\left|1\right\rangle^{(j)}\left|0\right\rangle^{(k)} \right).
\end{align}
which, in the instance of $J$ errors occurring in only the first redundant encoding, allows us to write 
\begin{widetext}
	\begin{align}
		\left\langle\mathbf{\epsilon}\right|_{\epsilon}^{1\rightarrow J} \left\langle\mathbf{0}\right|_{\epsilon}^{J+1\rightarrow n} \hat{U}\left|\psi\right\rangle_L
		=&\frac{\alpha}{\sqrt{2}}\left(\bigotimes^{k}\hat{u}_{T}\right)\left(\delta\Theta\left|0\right\rangle^{(k)}+\delta\Phi\left|1\right\rangle^{(k)}\right)\bigotimes^{q-1}\left[\left(\bigotimes^{n}\hat{u}_{T}\right)\left|0\right\rangle^{(n)}\right] \nonumber\\
		&+\frac{\beta}{\sqrt{2}}\left(\bigotimes^{k}\hat{u}_{T}\right)\left(\delta\Theta\left|1\right\rangle^{(k)}+\delta\Phi\left|0\right\rangle^{(k)} \right)\bigotimes^{q-1}\left[\left(\bigotimes^{n}\hat{u}_{T}\right)\left|1\right\rangle^{(n)}\right] \nonumber\\
		=&\frac{\alpha}{2}\left(\bigotimes^{k}\hat{u}_{T}\right)\left[\left(\delta\Theta+\delta\Phi\right)\left|+\right\rangle^{\otimes k}+\left(\delta\Theta - \delta\Phi\right)\left|-\right\rangle^{\otimes k}\right]\bigotimes^{q-1}\left[\left(\bigotimes^{n}\hat{u}_{T}\right)\left|0\right\rangle^{(n)}\right] \nonumber\\
		&+\frac{\beta}{2}\left(\bigotimes^{k}\hat{u}_{T}\right)\left[\left(\delta\Theta+\delta\Phi\right)\left|+\right\rangle^{\otimes k}-\left(\delta\Theta - \delta\Phi\right)\left|-\right\rangle^{\otimes k}\right]\bigotimes^{q-1}\left[\left(\bigotimes^{n}\hat{u}_{T}\right)\left|1\right\rangle^{(n)}\right]
	\end{align}
\end{widetext}
where $J+k=n$, $\left|\mathbf{0}\right\rangle_{\epsilon}^{a\rightarrow b}$ corresponds to error modes $a$ through to $b$ in the vacuum (no errors) while $\left|\epsilon\right\rangle_{\epsilon}^{a\rightarrow b}$ corresponds to an error occuring in modes $a$ through to $b$. Also 
\begin{align}
	\delta\Theta=&\left\langle\mathbf{\epsilon}\right|_{\epsilon}^{1\rightarrow J}\left(\bigotimes_{i=1}^{J}\hat{\mathcal{U}}_{i,e}\right)\left|0\right\rangle^{(J)} \\
	\textrm{and }\delta\Phi=&\left\langle\mathbf{\epsilon}\right|_{\epsilon}^{1\rightarrow J}\left(\bigotimes_{i=1}^{J}\hat{\mathcal{U}}_{i,e}\right)\left|1\right\rangle^{(J)}
\end{align}
are stochastic $\mathbb{C}$ numbers whose value depends on the specific location of the errors and the individual operations implemented on the qubits. After renormalisation, these factors randomly take the values $\pm1$.

If any one of the remaining $k$ physical qubits of the first redundant encoding are measured in the $\hat{u}_{T}\left|\pm\right\rangle^{(1)}=\frac{1}{\sqrt{2}}\hat{u}_{T}\left(\left|0\right\rangle^{(1)}\pm\left|1\right\rangle^{(1)}\right)$ basis, the state gains a corresponding and unimportant global phase factor of $\left(\delta\Theta+\delta\Phi\right)$ for $\left|+\right\rangle^{(1)}$ and $\left(\delta\Theta-\delta\Phi\right)$ for $\left|-\right\rangle^{(1)}$ and if $\left|-\right\rangle^{(1)}$ is returned there will also be a sign error on the logical $\left|1\right\rangle_{L}$ state. Thus, dropping the global phase, the final state after projective measurements on all error channels and a single un-errored channel, produces the final state after renormalisation
\begin{align}
	\left|\psi_{out}(\pm)\right\rangle_L=&\left(\bigotimes^{k-1}\hat{u}_{T}\left|\pm\right\rangle\right)\nonumber\\
	&\otimes\hat{U}_{T}\left(\alpha\bigotimes^{q-1}\left|0\right\rangle^{(n)}\pm\beta\bigotimes^{q-1}\left|1\right\rangle^{(n)}\right).
\end{align}
That is to say, the partially errored, redundent state is unentangled from the remainder of the state which successfully has the target unitary applied, although with a potential known phase error. This can be repeated for any of the remaining $q-1$ redundant copies in the parity encoding in which an error is detected. Provided at least one entire redundant copy is heralded as error free, the target unitary will be successfully applied, with a potential phase error if an odd number of $\left|-\right\rangle$ states are returned during the projective measurements. We therefore have the set of success criteria as
\begin{enumerate}
	\item At most $q-1$ redundant copies of the encoding are heralded to have encountered an error. This ensures there remains a logical state on which the gate is successfully applied.
	\item For all redundant copies of the encoding which are heralded to have encountered an error, at least one physical qubit must not have been heralded to be in error so that the projected measurement can be performed.
\end{enumerate}

\section{unitary averaging and Standard Fault Tolerance \label{sec:FT encodings}}
Now that we have established the action of UA on single and two cubit gates and considered how it can be integrated into existing loss correction schemes, it would be useful to attempt to translate these results into fault tolerant error correction. The purpose here is to explore the benefits of employing UA within existing error correction schemes to expand the parameter space for which fault tolerance can be achieved. That is, we will consider how the fault tolerant parameter space changes if the operations performed on each physical qubit are done utilising UA. Specifically we will concern ourselves with the per gate depolarisation probability $\varepsilon$  and per qubit, per gate loss rate $\gamma$. 

The easiest consideration is that of photon loss. This is because the UA scheme does nothing to correct for losses, but does increase the optical depth due to the encoding and decoding circuits. As discussed earlier, it will increase the optical depth per gate by $2\log_{2}\left(N\right)$. Furthermore, any heralded error will result in an effective loss, although a located one. Here, we will not seek to use the located nature of this loss to our advantage and treat it similarly to all other optical losses. As such, the UA scheme acts to increase the effective loss rate depending on the depth of encoding and decoding steps, and the probability of success. If the original loss rate was $\gamma$ per qubit per gate, then we can approximate the new effective loss rate for a gate as $\Gamma=\frac{\gamma}{3}\left(3+2\log_{2}\left(N\right)\right)+\left(1-P_{s}(N) \right)$ per qubit per gate. Here, we have assumed all gates to have an optical depth of a single qubit gate ($3$) and that encode and decode steps are as lossy as each individual component within the gate. Given some gates are likely to be much deeper than this we are likely over estimating the relative loss introduced by the additional encoding.

We next turn our attention to the error rates. To achieve this, consider that the gate fidelity relates to the probability that the qubit(s) it acts on would be measured in the incorrect state. After the gate is applied, with probability $\mathcal{F}(N)=1-\varepsilon$, the correct transformation is applied, and thus with probability $\varepsilon$ any syndrome measurement will herald an error. This suggests an equivalence between our characteristic noise parameter ($\mathcal{V}$) and the typical noise parameter considered in fault tolerance ($\varepsilon$) (the per qubit, per gate depolarisation rate). This typical noise parameter corresponds to the more general depolarisation errors occurring, which are not present in this model. However, if we take the depolarisation effects as sourced by these same stochastic noises, we can indeed treat these two noise terms as equivalent. Comparing to the fidelity shown in Equation \ref{eq:first order fidelity}, we have an effective gate error rate of $\mathcal{E}=\frac{\mathcal{\varepsilon}}{N+\varepsilon-N\varepsilon}$.

\begin{figure}
	\centering
     \begin{subfigure}{0.9\columnwidth}
		\includegraphics[width=\columnwidth]{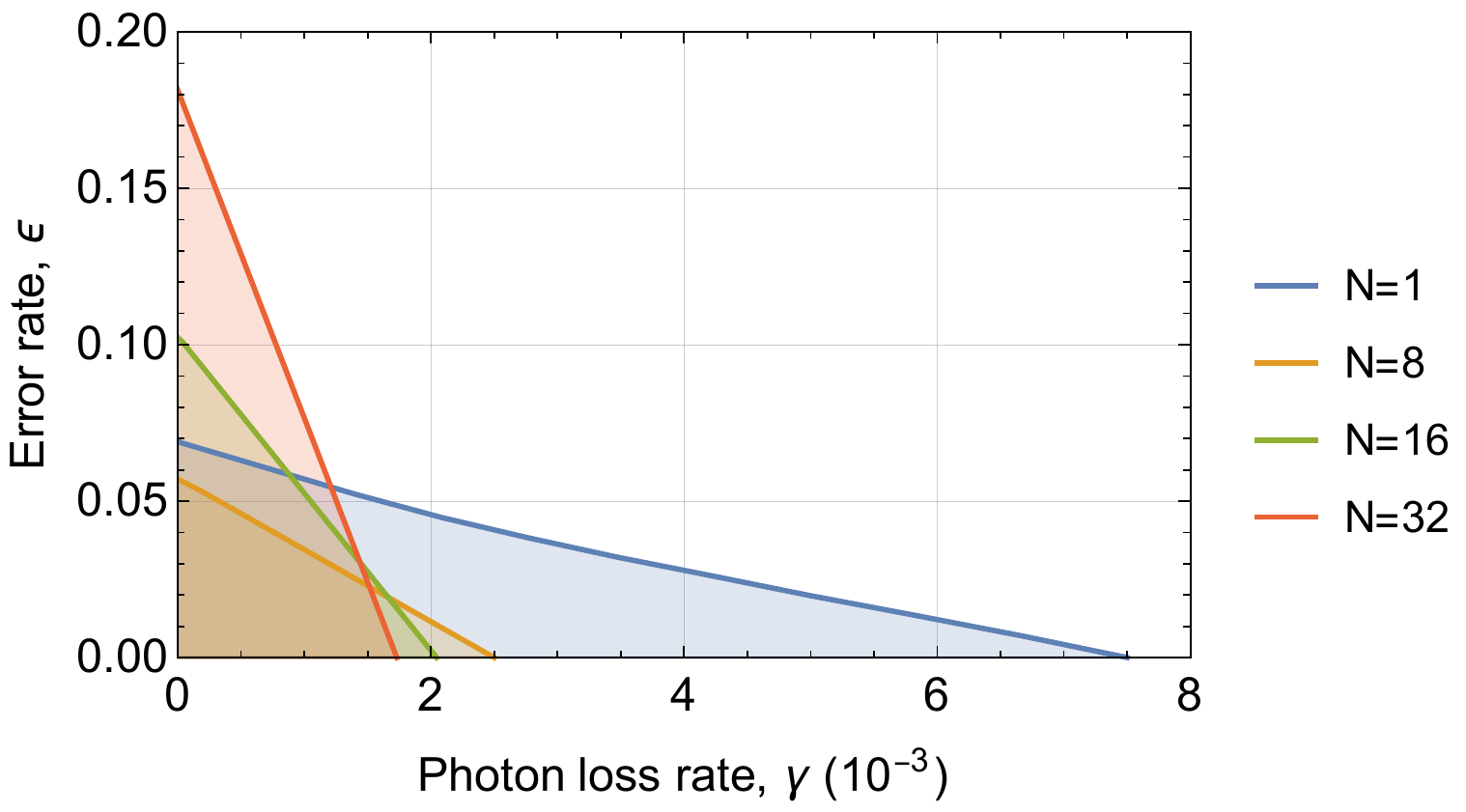}
		\caption{FBQC 4-star architecture.\label{fig:parameter space improvement 4 star}}
	\end{subfigure}
     \begin{subfigure}{0.9\columnwidth}
		\includegraphics[width=\columnwidth]{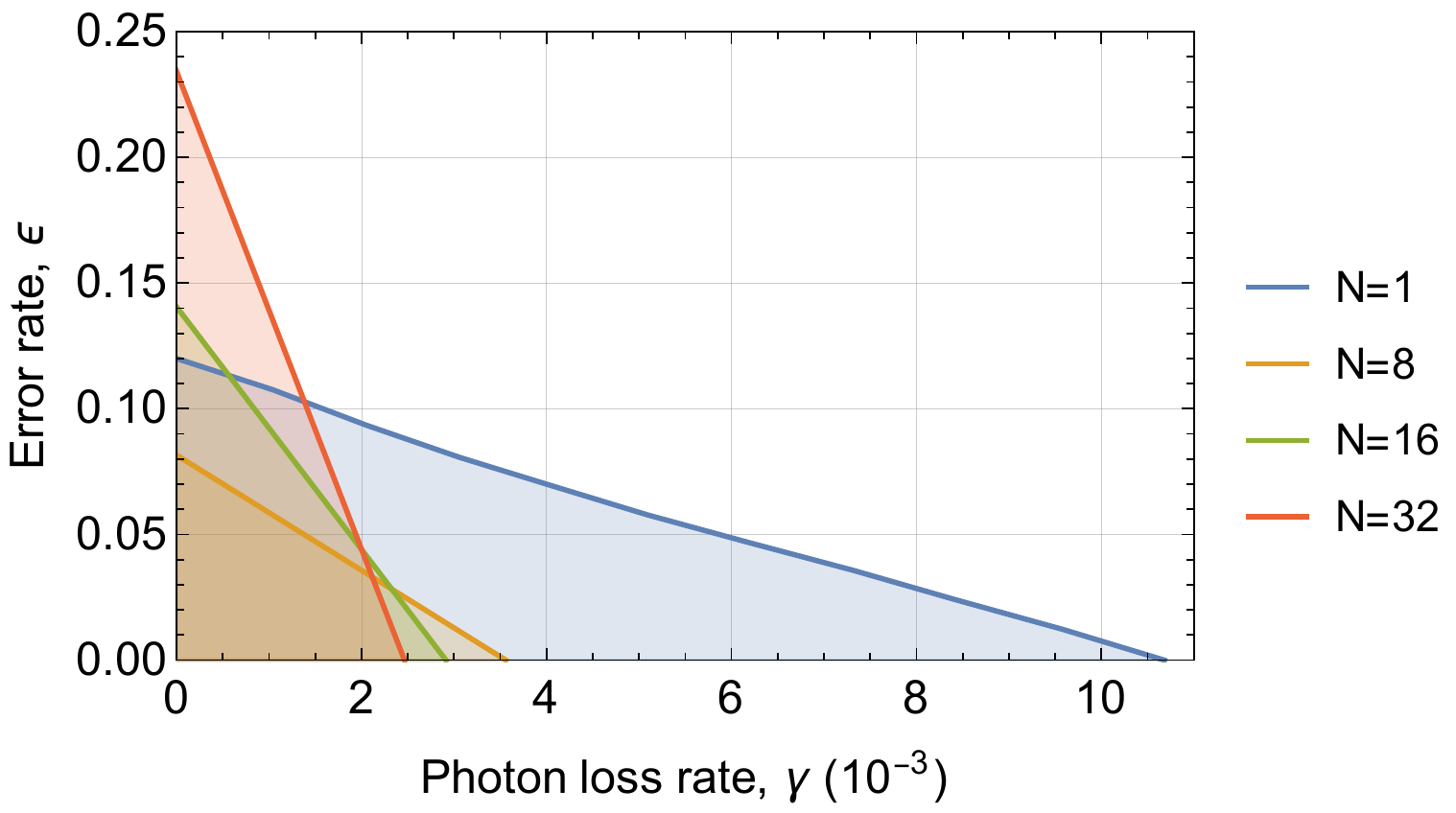}
		\caption{FBQC 6-ring architecture.\label{fig:parameter space improvement 6 ring}}
	\end{subfigure}
	\begin{subfigure}{0.9\columnwidth}
		\includegraphics[width=\columnwidth]{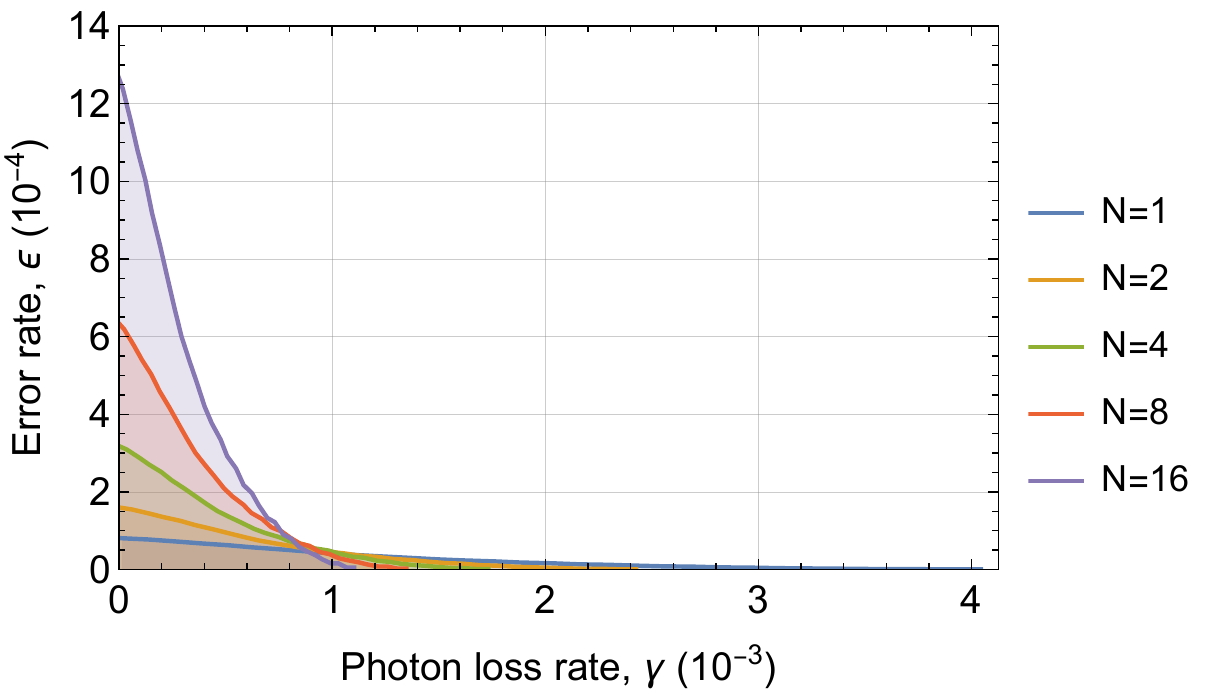}
		\caption{7-qubit Steane code.\label{fig:parameter space improvement b}}
	\end{subfigure}
	\begin{subfigure}{0.9\columnwidth}
		\includegraphics[width=\columnwidth]{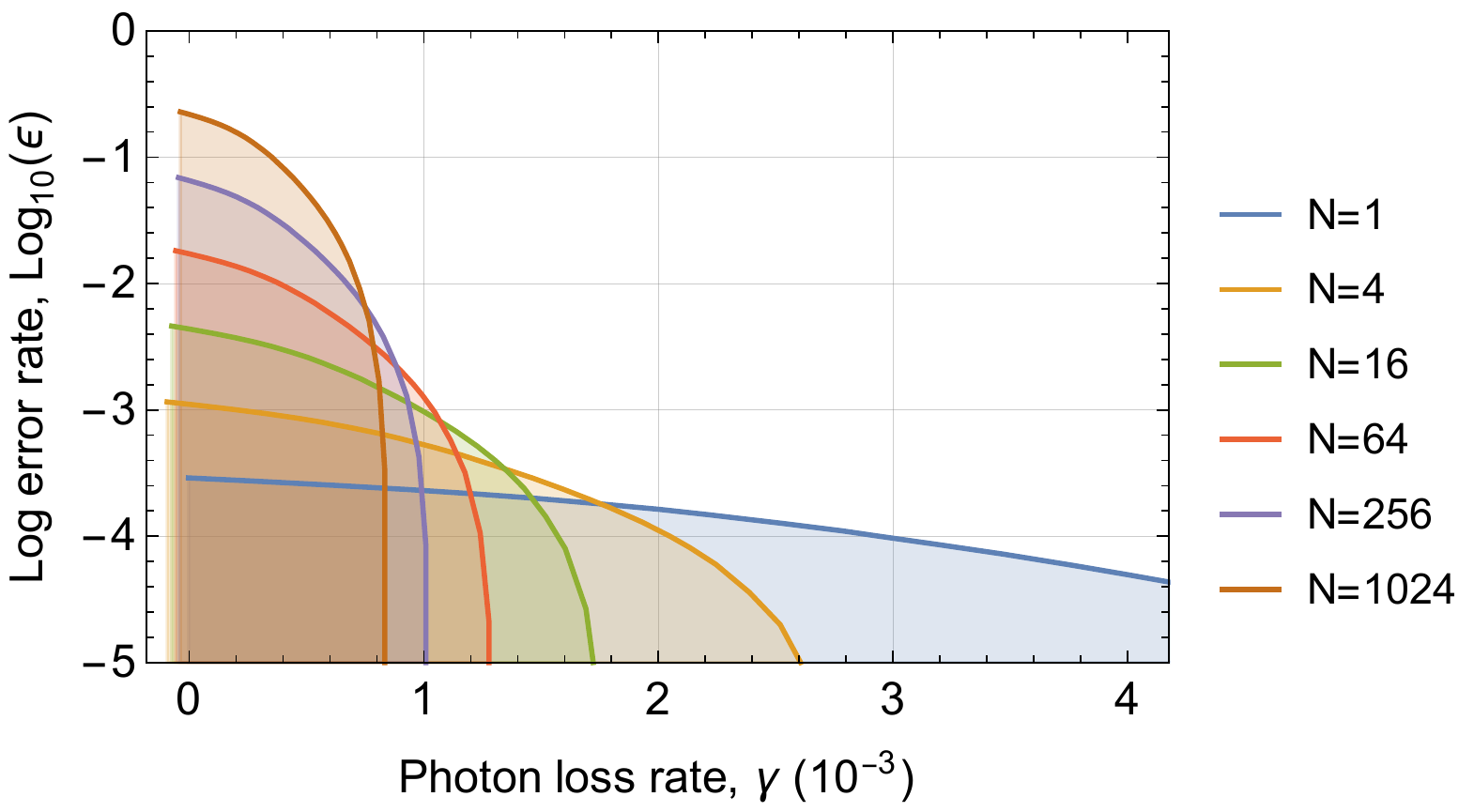}
		\caption{23-qubit Golay code. \label{fig:parameter space improvement c}}
	\end{subfigure}
	\caption{Fault tolerant parameter space improvement estimation when employing unitary averaging in a FBQC utilising a 4-star (\ref{fig:parameter space improvement 4 star}) and 6-ring (\ref{fig:parameter space improvement 6 ring}) architecture, data taken from Figure 3 in \cite{bartolucci2023fusion}, a 7-qubit Steane code (\ref{fig:parameter space improvement b}) and the 23-qubit Golay code (\ref{fig:parameter space improvement c}). $N=1$ data taken from \cite{dawson2006noise} for the surface code and the Steane and Golay codes respectively. Fault tolerance is achieved within the shaded region for each value of $N$. \label{fig:parameter space improvement} }
\end{figure}

In the same manner as we discussed above with the loss tolerant parity code, we can consider UA to sit within existing algorithms and serve to reduce the effective gate error at the cost of increasing loss. We will use the above results in conjunction with existing fault tolerant thresholds for some example error correction schemes. If we take the earlier results of Dawson, Haselgrove and Nielsen \cite{dawson2006noise} and Fujii and Tokunaga \cite{fujii2012error}, we can estimate the effect UA has on the fault tolerant parameter space for a %square lattice surface code, 
7-qubit Steane code and 23-qubit Golay code using Equations \ref{eq:first order fidelity} and \ref{eq:first order Psuccess}. We can also consider its effect on a more modern fault tolerant architecture such as Fusion-based Quantum Computation (FBQC) \cite{bartolucci2023fusion}. The results are shown in Figure \ref{fig:parameter space improvement}. This suggests a significant improvement in the parameter space at which fault tolerance can be achieved despite the increase in effective loss due to a lossy encoding circuit and the probabilistic nature of unitary averaging. This improvement primarily arises due to the difference in scaling between the effective loss rate $\Gamma$ and the gate error rate $\mathcal{E}$:
\begin{align}
	\Gamma=&\frac{\gamma}{3}\left(3+2\log_{2}\left(N\right)\right)+\varepsilon\left(1-\frac{1}{N}\right) \\
	\mathcal{E}=&\frac{\mathcal{\varepsilon}}{N+\varepsilon-N\varepsilon}
\end{align}
Specifically, for large $N$ we see loss scales logarithmically while the error scales as $\propto 1/N$. 

It is worth highlighting that the results in Figure \ref{fig:parameter space improvement} are dependent on some fairly large assumptions. Specifically that each component that goes into implementing these encodings and optical circuits can (1) be implemented using UA and (2) that UA has the same effect on their output as seen above. For example, implementing a measurement based quantum computer will likely require many components not considered above such as quantum memories and detectors. It seems reasonable that memories will be compatible with UA, however destructive measurements can clearly not be treated as unitaries which can be averaged, thus these results will likely only apply to the preparation stages of such a device. Specifically considering the scaling shown in Figures \ref{fig:parameter space improvement 4 star} and \ref{fig:parameter space improvement 6 ring}, the $N=1$ results are for an implementation agnostic analysis for FBQC. For an optical implementation, at least some of the photon loss will be due to the probabilistic nature of fusion gates and imperfect detectors. We are here treating all loss as due to photon absorption in this situation, and so are overestimating the additional loss induced with the increased optical depth of UA. This has the effect of underestimating the tolerable loss to maintain fault tolerance. As such, these results need to be viewed as approximate scaling.

We also see a perhaps surprising result in Figures \ref{fig:parameter space improvement 4 star} and \ref{fig:parameter space improvement 6 ring}, where a benifit is only observed for sufficiently large $N$, When $N\le8$, the additional induced loss eliminates any benefit of improved tolerable error rates, as a result, only with moderate large $N$ can any benefit from applying UA be gained.

\section{Conclusion \label{sec:Conclusion}}

This paper has explored the use of UA within the context of applying arbitrary one and two qubit gate transformations. We have shown that UA can be implemented in such a way as to avoid encoding errors while also introducing only a logarithmic number of optical elements to the path depth. Unitary averaging may enable arbitrary high gate fidelity while at the fixed cost of a reduced probability of success which is linear in the initial parameter variance within a single unitary. We have also demonstrated that this loss can be mitigated by employing existing techniques for loss protection which will already be necessary in any large scale system. Furthermore, we have shown that UA can be used within existing fault tolerant schemes to modify and potentially improve the parameter space for which error tolerance may be achieved. We have done so considering both early fault tolerant schemes which can greatly benefit from the use of UA as well as more modern, surface code based schemes for which the benefit of applying UA appears to be more modest.

\section*{Acknowledgements}
 APL acknowledges support from BMBF (QPIC) and the Einstein Research Unit on Quantum Devices. This research was supported by the
Australian Research Council (ARC) under the Centre of Excellence for Quantum Computation and Communica-
tion Technology (Project No. CE170100012).

\bibliography{bibliography}
\begin{widetext}
\appendix
\section{Single Qubit Gate Errors Calculation \label{app:Single qubit gate errors}}
Here I present the details of so solving for the expectation values used to calculate the probability of success when error averaging a single qubit gate.
\subsection{Errored Single Qubit Gate \label{app:errored single qubit gate}}
Here we consider each physical gate to be averaged $\hat{U}_{j}$. Taylor expanding each parameter and keeping only the terms which are bilinear or quadratic in the error terms gives
\begin{align}
	\hat{U}_{j}=&\begin{bmatrix}
		e^{i\phi_{1,j}}e^{i\chi_{1,j}}\sin\left(\theta_{j}\right) & e^{i\phi_{2,j}}e^{i\chi_{1,j}}\cos\left(\theta_{j}\right) \\
		e^{i\phi_{1,j}}e^{i\chi_{2,j}}\cos\left(\theta_{j}\right) & -e^{i\phi_{2,j}}e^{i\chi_{2,j}}\sin\left(\theta_{j}\right)
	\end{bmatrix} \nonumber\\
	=&\begin{bmatrix}
		a_{j} & b_{j} \\
		c_{j} & d_{j}
	\end{bmatrix}
\end{align}
where
\begin{align}
	a_{j}=&e^{i\phi_{1}}e^{i\chi_{1}}\sin\left(\theta\right)+\Bigg[e^{i\phi_{1}}e^{i\chi_{1}}\sin\left(\theta\right)\left(i\delta\phi_{1,j}+i\delta\chi_{1,j}-\delta\phi_{1,j}\delta\chi_{1}-\frac{\delta\phi_{1,j}^2}{2}-\frac{\delta\chi_{1,j}^2}{2}-\frac{\delta\theta_{j}^2}{2}\right) \nonumber\\
	&+e^{i\phi_{1}}e^{i\chi_{1}}\cos\left(\theta\right)\delta\theta_{j}\left(1+i\delta\phi_{1,j}+i\delta\chi_{1,j}\right)\Bigg] \equiv a+\left[\delta a_{j}\right] \\
	b_{j}=&e^{i\phi_{2}}e^{i\chi_{1}}\cos\left(\theta\right)+\Bigg[e^{i\phi_{2}}e^{i\chi_{1}}\cos\left(\theta\right)\left(i\delta\phi_{2,j}+i\delta\chi_{1,j}-\delta\phi_{2,j}\delta\chi_{1,j}-\frac{\delta\phi_{2,j}^2}{2}-\frac{\delta\chi_{1,j}^2}{2}-\frac{\delta\theta_{j}^2}{2}\right)  \nonumber\\
	&-e^{i\phi_{2}}e^{i\chi_{1}}\sin\left(\theta\right)\delta\theta_{j}\left(1+i\delta\phi_{2,j}+i\delta\chi_{1,j}\right)\Bigg] \equiv b+\left[\delta b_{j}\right]\\
	c_{j}=&e^{i\phi_{1}}e^{i\chi_{2}}\cos\left(\theta\right)+\Bigg[e^{i\phi_{1}}e^{i\chi_{2}}\cos\left(\theta\right)\left(i\delta\phi_{1,j}+i\delta\chi_{2,j}-\delta\phi_{1,j}\delta\chi_{2,j}-\frac{\delta\phi_{1,j}^2}{2}-\frac{\delta\chi_{2,j}^2}{2}-\frac{\delta\theta_{j}^2}{2}\right)  \nonumber\\
	&-e^{i\phi_{1}}e^{i\chi_{2}}\sin\left(\theta\right)\delta\theta_{j}\left(1+i\delta\phi_{1,j}+i\delta\chi_{2,j}\right)\Bigg] \equiv c+\left[\delta c_{j}\right]\\
	d_{j}=&-e^{i\phi_{2}}e^{i\chi_{2}}\sin\left(\theta\right)+\Bigg[-e^{i\phi_{2}}e^{i\chi_{2}}\sin\left(\theta\right)\left(i\delta\phi_{2,j}+i\delta\chi_{2,j}-\delta\phi_{2,j}\delta\chi_{2,j}-\frac{\delta\phi_{2,j}^2}{2}-\frac{\delta\chi_{2,j}^2}{2}-\frac{\delta\theta_{j}^2}{2}\right) \nonumber\\
	&-e^{i\phi_{2}}e^{i\chi_{2}}\cos\left(\theta\right)\delta\theta_{j}\left(1+i\delta\phi_{2,j}+i\delta\chi_{2,j}\right)\Bigg]  \equiv d+\left[\delta d_{j}\right].
\end{align}
Each term within the square bracket represents the noise term unique to each physical copy of the unitary while the preceding term is the intended value of each matrix element. Thus we can write
\begin{equation}
	\hat{U}_{j}=\begin{bmatrix}
		a & b \\
		c & d
	\end{bmatrix} + \begin{bmatrix}
		\delta a_{j} & \delta b_{j} \\
		\delta c_{j} & \delta d_{j}
	\end{bmatrix}=\hat{U}_{T}+\hat{E}_{j}.
\end{equation}
\subsection{Solving For Probability of Success \label{app:Psuccess}}
To solve for the probability of success $P_s(N)$ we assume each noise term is independent, with $\left\langle\delta O\right\rangle=0$ and $\left\langle\delta O^2\right\rangle=\nu$ for all terms $\delta O$ and use the Taylor expanded form of the applied transformations as given above. We also consider the general input state 
\begin{equation}
	\left|\psi\right\rangle=\begin{bmatrix}
		\alpha \\
		\beta \end{bmatrix}.
\end{equation} 
The probability of success is then given by
\begin{align}
	P_{s}(N)=&\left\langle \psi\right| \hat{\mathcal{U}}^{\dagger}(N)\hat{\mathcal{U}}(N)\left|\psi\right\rangle\nonumber\\
	=&\frac{1}{N^2}\sum_{j=1}^{N}\sum_{k=1}^{N}\left\langle \psi\right| \hat{U}_{j}^{\dagger}\hat{U}_{k}\left|\psi\right\rangle \nonumber\\
	=&\frac{1}{N^2}\sum_{j=1}^{N}\sum_{k=1}^{N}\Bigg[\left|\alpha\right|^2\left\langle a_j^{*}a_{k}+c_{j}^{*}c_{k}\right\rangle \nonumber\\
	&+\left|\beta\right|^2\left\langle b_j^{*}b_{k}+d_{j}^{*}d_{k}\right\rangle+\alpha^{*}\beta\left\langle a_j^{*}b_{k}+c_{j}^{*}d_{k}\right\rangle\nonumber\\
	&+\alpha\beta^{*}\left\langle b_j^{*}a_{k}+d_{j}^{*}c_{k}\right\rangle\Bigg]
\end{align}
Going term-by-term through this gives
\begin{align}
	\frac{1}{N^2}\sum_{j=1}^{N}\sum_{k=1}^{N}\left\langle a_{j}^{*}a_{k}\right\rangle=&\frac{1}{N^2}\sum_{j=1}^{N}\sum_{k=1}^{N}\Bigg\{\sin^2\left(\theta\right)\bigg[\left\langle1-\frac{1}{2}\left(\delta\phi_{1,j}^2+\delta\chi_{1,j}+\delta\theta_{1,j}^2\right)-\frac{1}{2}\left(\delta\phi_{1,k}^2+\delta\chi_{1,k}+\delta\theta_{1,k}^2\right)\right\rangle\nonumber\\
	&+\left\langle\frac{1}{4}\left(\delta\phi_{1,j}^2+\delta\chi_{1,j}^2+\delta\theta_{1,j}^2\right)\left(\delta\phi_{1,k}^2+\delta\chi_{1,k}^2+\delta\theta_{1,k}^2\right)\right\rangle+\delta_{j,k}\left\langle\delta\phi_{1,j}^2+\delta\chi_{1,j}^2+\delta\phi_{j}^2\delta\chi_{1,j}^2\right\rangle\bigg]\nonumber\\
	&+\cos^2\left(\theta\right)\delta_{j,k}\left\langle\theta_{j}^2\right\rangle\left\langle1+\delta\phi_{j}^2+\delta\chi_{j}^2\right\rangle\Bigg\} \nonumber\\
	=&\sin^2\left(\theta\right)\left(1-3\nu+\frac{9}{4}\nu^2+\frac{2\nu+\nu^2}{N}\right) + \cos^2\left(\theta\right)\frac{\nu+2\nu^2}{N}=\frac{1}{N^2}\sum_{j=1}^{N}\sum_{k=1}^{N}\left\langle d_{j}^{*}d_{k}\right\rangle,
\end{align}
\begin{equation}
	\frac{1}{N^2}\sum_{j=1}^{N}\sum_{k=1}^{N}\left\langle 	b_{j}^{*}b_{k}\right\rangle=\frac{1}{N^2}\sum_{j=1}^{N}\sum_{k=1}^{N}\left\langle c_{j}^{*}c_{k}\right\rangle=\cos^2\left(\theta\right)\left(1-3\nu+\frac{9}{4}\nu^2+\frac{2\nu+\nu^2}{N}\right) + \sin^2\left(\theta\right)\frac{\nu+2\nu^2}{N}
\end{equation}
and
\begin{align}
	\frac{1}{N^2}\sum_{j=1}^{N}\sum_{k=1}^{N}\left\langle 	a_{j}^{*}b_{k}\right\rangle=&\frac{e^{i\left(\phi_{2}-\phi_{1}\right)}}{N^2}\sum_{j=1}^{N}\sum_{k=1}^{N}\Bigg\{\sin\left(\theta\right)\cos\left(\theta\right)\bigg[\left\langle1-\frac{1}{2}\left(\delta\phi_{1,j}^2+\delta\chi_{1,j}+\delta\theta_{1,j}^2\right)-\frac{1}{2}\left(\delta\phi_{1,k}^2+\delta\chi_{1,k}+\delta\theta_{1,k}^2\right)\right\rangle\nonumber\\
	&+\left\langle\frac{1}{4}\left(\delta\phi_{1,j}^2+\delta\chi_{1,j}^2+\delta\theta_{1,j}^2\right)\left(\delta\phi_{1,k}^2+\delta\chi_{1,k}^2+\delta\theta_{1,k}^2\right)\right\rangle+\delta_{j,k}\left\langle\delta\phi_{1,j}^2+\delta\chi_{1,j}^2+\delta\phi_{j}^2\delta\chi_{1,j}^2\right\rangle\bigg]\nonumber\\
	&-\sin\left(\theta\right)\cos\left(\theta\right)\delta_{j,k}\left\langle\theta_{j}^2\right\rangle\left\langle1+\delta\phi_{j}^2+\delta\chi_{j}^2\right\rangle\Bigg\} \nonumber\\
	\implies&\frac{1}{N^2}\sum_{j=1}^{N}\sum_{k=1}^{N}\alpha^{*}\beta\left\langle a_{j}^{*}b_{k}\right\rangle + \alpha\beta^{*}\left\langle a_{j}b_{k}^{*}
	\right\rangle=2\left|\alpha\right|\left|\beta\right|\cos\left(\gamma\right)\sin\left(\theta\right)\cos\left(\theta\right)\left(1-3\nu+\frac{9}{4}\nu^2+\frac{\nu-\nu^2}{N}\right)
\end{align}
where $\gamma=\phi_{1}-\phi_{2}-\theta_{\alpha}+\theta_{\beta}$ for $\alpha=\left|\alpha\right|e^{i\theta_{\alpha}}$ and $\beta=\left|\beta\right|e^{i\theta_{\beta}}$. Similarly
\begin{align}
	\frac{1}{N^2}\sum_{j=1}^{N}\sum_{k=1}^{N}\left\langle 	c_{j}^{*}d_{k}\right\rangle=&-\frac{e^{i\left(\phi_{2}-\phi_{1}\right)}}{N^2}\sum_{j=1}^{N}\sum_{k=1}^{N}\Bigg\{\sin\left(\theta\right)\cos\left(\theta\right)\bigg[\left\langle1-\frac{1}{2}\left(\delta\phi_{1,j}^2+\delta\chi_{1,j}+\delta\theta_{1,j}^2\right)-\frac{1}{2}\left(\delta\phi_{1,k}^2+\delta\chi_{1,k}+\delta\theta_{1,k}^2\right)\right\rangle\nonumber\\
	&+\left\langle\frac{1}{4}\left(\delta\phi_{1,j}^2+\delta\chi_{1,j}^2+\delta\theta_{1,j}^2\right)\left(\delta\phi_{1,k}^2+\delta\chi_{1,k}^2+\delta\theta_{1,k}^2\right)\right\rangle+\delta_{j,k}\left\langle\delta\phi_{1,j}^2+\delta\chi_{1,j}^2+\delta\phi_{j}^2\delta\chi_{1,j}^2\right\rangle\bigg]\nonumber\\
	&-\sin\left(\theta\right)\cos\left(\theta\right)\delta_{j,k}\left\langle\theta_{j}^2\right\rangle\left\langle1+\delta\phi_{j}^2+\delta\chi_{j}^2\right\rangle\Bigg\} \nonumber\\
	\implies&\frac{1}{N^2}\sum_{j=1}^{N}\sum_{k=1}^{N}\alpha^{*}\beta\left\langle c_{j}^{*}d_{k}\right\rangle + \alpha\beta^{*}\left\langle d_{j}c_{k}^{*}
	\right\rangle=-2\left|\alpha\right|\left|\beta\right|\cos\left(\gamma\right)\sin\left(\theta\right)\cos\left(\theta\right)\left(1-3\nu+\frac{9}{4}\nu^2+\frac{\nu-\nu^2}{N}\right).
\end{align}
Putting this all together yields
\begin{equation}
	P_{s}\left(N\right)=1-3\nu+\frac{3\nu}{N}+\frac{9\nu^2}{4}+\frac{3\nu^2}{N}
\end{equation}
Note that this function is only accurate up to $\mathcal{O}(\nu)$ as $\nu^2$ terms can arise due to $\left\langle \mathcal{O}(\delta x^4)\mathcal{O}(\delta x^0)\right\rangle$ which have not been included. For this reason a calculation including fourth order in $\delta x$ was completed as above to yield 
\begin{align}
	P_{s}\left(N\right)\approx&1-3\nu+\frac{3\nu}{N}+4\nu^2-\frac{4\nu^2}{N} \nonumber\\
	&-\frac{21\nu^3}{8}+\frac{\nu^3}{12N}+\frac{49\nu^4}{64}+\frac{13\nu^4}{6N}
\end{align}
which is similarly only complete up to $\mathcal{O}(\nu^2)$. This is the cause of the erroneous behaviour where $P_{s}\left(N=1\right)<0$ for $\nu>0$.

\subsection{Solving For Fidelity\label{app:Fidelity}}
To solve for the fidelity can be done in much the same manner. Particularly as after re-normalisation the corrected state will be pure. As such we can write the corrected state fidelity as 
\begin{align}
	\mathcal{F}(N)=&\left\langle \Psi\right| \hat{\rho}_{ps}(N)\left|\Psi\right\rangle\nonumber\\
	=&\left(P_s(N)\right)^{-1}\left(1+\frac{1}{N}\sum_{j=1}^{N}\left\langle \psi\right| \hat{U}_{T}^{\dagger}\hat{E}_{j}\left|\psi\right\rangle\right)^2 \nonumber\\
	=&\left(P_s(N)\right)^{-1}\Bigg(1+\frac{\mathcal{N}}{N}\sum_{j=1}^{N}\bigg[\left|\alpha\right|^2\left\langle a^{*}\delta a_{j}+c^{*}\delta c_{j}\right\rangle +\left|\beta\right|^2\left\langle b^{*}\delta b_{j}+d^{*}\delta d_{j}\right\rangle+\alpha^{*}\beta\left\langle a^{*}\delta b_{j}+c^{*}\delta d_{j}\right\rangle\nonumber\\
	&+\alpha\beta^{*}\left\langle b^{*}\delta a_{j}+d^{*}\delta c_{j}\right\rangle\bigg]\Bigg)
\end{align}
So once again going term-by-term, and here expanding $\delta a_{j}/\delta b_{j}/\delta c_{j}/\delta d_{j}$ up to fourth order in the error terms give
\begin{equation}
	\frac{1}{N}\sum_{j=1}^{N}\left\langle a^{*}\delta a_{j}\right\rangle=\frac{1}{N}\sum_{j=1}^{N}\left\langle d^{*}\delta d_{j}\right\rangle=\sin^2\left(\theta\right)\left(-\frac{3\nu}{2}+\frac{7\nu^2}{8}\right),
\end{equation}
\begin{equation}
	\frac{1}{N}\sum_{j=1}^{N}\left\langle b^{*}\delta b_{j}\right\rangle=\frac{1}{N}\sum_{j=1}^{N}\left\langle c^{*}\delta c_{j}\right\rangle=\cos^2\left(\theta\right)\left(-\frac{3\nu}{2}+\frac{7\nu^2}{8}\right),
\end{equation}
and
\begin{align}
	\frac{1}{N}\sum_{j=1}^{N} \alpha^{*}\beta \left\langle a^{*}\delta b_{j}\right\rangle+\alpha\beta^{*} \left\langle b^{*}\delta a_{j}\right\rangle=&2\left|\alpha\right|\left|\beta\right|\cos\left(\gamma\right)\sin\left(\theta\right)\cos\left(\theta\right)\left(-\frac{3\nu}{2}+\frac{7\nu^2}{8}\right)
\end{align}
where $\gamma=\phi_{1}-\phi_{2}-\theta_{\alpha}+\theta_{\beta}$ for $\alpha=\left|\alpha\right|e^{i\theta_{\alpha}}$ and $\beta=\left|\beta\right|e^{i\theta_{\beta}}$. Similarly
\begin{align}
	\frac{1}{N}\sum_{j=1}^{N} \alpha^{*}\beta\left\langle c^{*}\delta d_{j}\right\rangle + \alpha\beta^{*}\left\langle d^{*}\delta c_{j}
	\right\rangle=&-2\left|\alpha\right|\left|\beta\right|\cos\left(\gamma\right)\sin\left(\theta\right)\cos\left(\theta\right)\left(-\frac{3\nu}{2}+\frac{7\nu^2}{8}\right)
\end{align}
Putting this all together yields
\begin{align}
	\mathcal{F}\left(N\right)\approx&\left(P_{s}(N)\right)^{-1}\left(1-\frac{3\nu}{2}+\frac{7\nu^2}{8}\right)^2 \nonumber\\
	\approx&\left(1-3\nu+\frac{3\nu}{N}+4\nu^2-\frac{4\nu^2}{N}\right)^{-1}\left(1-3\nu+4\nu^2\right)
\end{align}

\section{Fusion Gates}
The Type-II gate is implemented by the transformation
\begin{align}
	F_{II}=&\begin{bmatrix}
		\sin\left(\theta_3\right) & \cos\left(\theta_3\right) & 0 & 0 \\
		\cos\left(\theta_3\right) & -\sin\left(\theta_3\right) & 0 & 0 \\
		0 & 0 & \sin\left(\theta_4\right) & \cos\left(\theta_4\right) \\
		0 & 0 & \cos\left(\theta_4\right) & -\sin\left(\theta_4\right)
	\end{bmatrix} \begin{bmatrix}
		1 & 0 & 0 & 0 \\
		0 & 0 & 0 & 1 \\
		0 & 0 & 1 & 0 \\
		0 & 1 & 0 & 0
	\end{bmatrix} \begin{bmatrix}
		\sin\left(\theta_1\right) & \cos\left(\theta_1\right) & 0 & 0 \\
		\cos\left(\theta_1\right) & -\sin\left(\theta_1\right) & 0 & 0 \\
		0 & 0 & \sin\left(\theta_2\right) & \cos\left(\theta_2\right) \\
		0 & 0 & \cos\left(\theta_2\right) & -\sin\left(\theta_2\right)
	\end{bmatrix}  \nonumber\\
	=&\begin{bmatrix}
		\sin\left(\theta_1\right)\sin\left(\theta_3\right) & \cos\left(\theta_1\right)\sin\left(\theta_3\right) & \cos\left(\theta_2\right)\cos\left(\theta_3\right) & -\sin\left(\theta_2\right)\cos\left(\theta_3\right) \\
		\sin\left(\theta_1\right)\cos\left(\theta_3\right) & \cos\left(\theta_1\right)\cos\left(\theta_3\right) & -\cos\left(\theta_2\right)\sin\left(\theta_3\right) & \sin\left(\theta_2\right)\sin\left(\theta_3\right) \\
		\cos\left(\theta_1\right)\cos\left(\theta_4\right) & -\sin\left(\theta_1\right)\cos\left(\theta_4\right) & \sin\left(\theta_2\right)\sin\left(\theta_4\right) & \cos\left(\theta_2\right)\sin\left(\theta_4\right) \\
		-\cos\left(\theta_1\right)\sin\left(\theta_4\right) & \sin\left(\theta_1\right)\sin\left(\theta_4\right) & \sin\left(\theta_2\right)\cos\left(\theta_4\right) & \cos\left(\theta_2\right)\cos\left(\theta_4\right) \\
	\end{bmatrix}
\end{align}

Preceeding as before, but with the aid of \emph{Mathematica}, we can see for a general two photon input state the probability of success is given by
\begin{equation}
	P_{\textrm{s, II}}(N)=1-2\nu+\frac{2\nu}{N}+\frac{5\nu^2}{3}-\frac{5\nu^2}{3N}
\end{equation}
and the post selected fidelity of
\begin{equation}
	\mathcal{F}_{\textrm{II}}(N)=P_{\textrm{s, II}}(N)^{-1}\left(1-2\nu+\frac{5\nu^2}{3}\right)
\end{equation}
where once again, each parameter is taken to have an equal but independent noise spectrum with the parameter $O=O_{T}+\delta O$, $\left\langle O\right\rangle=O_{T}$, $\left\langle\delta O\right\rangle=0$, $\left\langle\delta O^2\right\rangle=\nu$, $\left\langle\delta O^3\right\rangle=0$, $\left\langle\delta O^4\right\rangle=\nu^2$.
\end{widetext}
\end{document}